\pdfoutput=1
\documentclass[aps,prd,superscriptaddress,floatfix,nofootinbib,preprintnumbers,eqsecnum,twocolumn]{revtex4-1}

\usepackage{graphicx}
\usepackage{dcolumn}
\usepackage{bm}
\usepackage{comment}
\usepackage{xcolor}
\usepackage{amsmath}
\usepackage[colorlinks=true,citecolor=blue]{hyperref}
\usepackage{soul}
\usepackage{orcidlink}

\usepackage[normalem]{ulem}
\usepackage[dvipsnames]{xcolor}

\begin{document}

\title{On Stochastic Memory Backgrounds In LISA}

\author{James Buda\,\orcidlink{0009-0000-7939-5989}}
\email{james.buda@stonybrook.edu}
\affiliation{Department of Physics and Astronomy, University of California, Irvine, Irvine, CA 92617, USA}
\affiliation{Department of Physics and Astronomy, Stony Brook University, Stony Brook, NY 11794, USA}
\author{Andrew Laeuger\,\orcidlink{0000-0002-8212-6496}}
\author{Yanbei Chen\,\orcidlink{0000-0002-9730-9463}}
\affiliation{Theoretical Astrophysics, California Institute of Technology, Pasadena, CA 91125, USA}

\date{\today}

\begin{abstract}
Gravitational-wave (GW) memory is a permanent change in spacetime geometry induced by a burst of gravitational waves. It is theoretically interesting for its connection to asymptotic symmetries, but has not yet been observed. 
The upcoming Laser Interferometer Space Antenna (LISA) is predicted to detect memory directly in loud events from supermassive black hole (SMBH) mergers. We assess the prospects for the cumulative effect of memory contributions as a stochastic GW memory background (SGWMB). Using two SMBH population models, we recover previously obtained single-event SNR forecasts and compute the 
power spectrum of the unresolved background, using numerical-relativity waveforms for modeling the memory signal during each event. 
We find the predicted spectrum is significantly smaller than ones found by modeling the memory waveform with a Heaviside function, with 
expected SNRs of order $\mathcal{O}(0.1-10)$. 
We further find that our models favor a highly non-Gaussian (``popcorn") spectrum, demonstrating that the background may not appear as the union of continuous, overlapping signals, but rather as a collection of intermittent bursts. In both measures, we demonstrate that any current prediction of the spectrum is heavily dependent on the choice of population model. 
We interpret these findings in the light of the future LISA mission and emphasize their importance for properly handling memory within the LISA global fit. 
\end{abstract}

\maketitle

\section{Introduction}\label{sec:intro}

The study of gravitational waves and their physical discovery with ground-based gravitational wave detectors, including LIGO, VIRGO and KAGRA~\cite{LIGOScientific:2016aoc, LIGOScientific:2018mvr, LIGOScientific:2020ibl, KAGRA:2021vkt, LIGOScientific:2025slb, LIGO_web, VIRGO_web, KAGRA_web},
has provided profound insight into the theory of General Relativity~\cite{LIGOScientific:2016lio, LIGOScientific:2018dkp, LIGOScientific:2020tif, LIGOScientific:2021sio}
and astrophysical phenomena regarding black holes~\cite{LIGOScientific:2020kqk, KAGRA:2021duu}. Gravitational waves emitted by the coalescence of binary black holes mainly consist of the inspiral, merger and ringdown parts --- which are broadly oscillatory, due to the orbital motion of the two individual progenitor black holes at the beginning and the settling down of the final remnant black hole at the end.  However, there does exist a {\it non-oscillatory} piece of the gravitational-wave signal from a compact binary merger that gradually rises up, with unequal initial and final values, associated with the so-called memory effect~\cite{Favata:2009ii,Favata:2009ii}.
This permanent change in spacetime geometry can be registered locally by gravitational-wave detectors~\cite{braginsky1987gravitational}, separates into linear and nonlinear memories~\cite{Christodoulou:1991,Thorne:1992PRD45_520}, and can be connected to the soft-graviton theorem \cite{PhysRev.140.B516}. 

In gravitational wave detectors, the memory strain arises as an integrated, nonlinear response to the flux radiated during the coalescence rather than from the oscillatory dynamics themselves. Due to this its amplitude is generally much smaller than that of the inspiral-merger-ringdown signal it accompanies.
This smallness, together with its relatively long ramp-up time, makes memory waves considerably harder to detect than the oscillatory part of the signal. 
On the ground, the combined statistics from multiple events may allow detection of memory waves within the current generation~\cite{lasky2016detecting,Thrane:2013kb, Mitman:2026zfg}, while individual events may only be detectable in next-generation detectors~\cite{goncharov2024inferring,Grant:2022bla}.

Space-based detectors such as the Laser Interferometer Space Antenna (LISA)~\cite{Robson:2019qvq, Amaro-Seoane:2017uwf}, which will operate within the frequency band of $10^{-5}$ -- $10^{-1}\,\mathrm{Hz}$, offer much higher Signal-to-Noise Ratio (SNR) for Supermassive black-hole 
(SMBH) mergers, opening the door to memory detection where ground-based detectors fall short. The prospect of detecting memory contributions from SMBH mergers with LISA specifically has been explored by \cite{Favata:2009ii, Inchauspe:2024, Mitman:2024, Yang:2018, Pollney:2010}. 
Still, whereas the primary signal of massive black hole mergers in LISA tends to be high SNR and individually resolvable, most memory signals will not be resolvable by themselves \cite{Inchauspe:2024}. This leads to the realization that a \textit{stochastic background} may be built from unresolved memory events present within the detector \cite{Zhao:2021zlr}. 
Such a prospect is far from hypothetical: at nanohertz frequencies, pulsar timing array collaborations have recently reported evidence for a Hellings--Downs-correlated stochastic background~\cite{NANOGrav:2023gor, EPTA:2023fyk, Zic:2023gta, Xu:2023wog}.

Despite this, our understanding of the properties of this memory contribution remains limited, owing largely to how poorly constrained the SMBH population itself is. Two distinct models have been studied for the Poissonian arrivals of signals: (i) individually detectable events that are well separated in time-frequency space, and (ii) individually undetectable events that overlap substantially. Case (i) describes the detection of most SMBH mergers in LISA \cite{Klein:2015hvg, amaroseoane2017laserinterferometerspaceantenna}, while (ii) will lead to the formation of a SGWB, as is the case of some galactic and extragalactic neutron-star and/or white-dwarf binaries \cite{Nelemans:2001hp, Farmer:2003pa, LIGOScientific:2018dkp, Hils:1990vc}. Nevertheless, the boundary between a preferred description in terms of individual sources and a stochastic background can be fuzzy, with non-Gaussian stochastic backgrounds potentially capturing a ``popcorn'' regime of discrete sub-threshold bursts \cite{Thrane:2013kb,Cornish:2015pda}.

As it turns out, the individually undetectable portion of memory waves in LISA likely will not fit case (ii), as they do not appear frequently enough to substantially overlap with each other. 
Here it is useful to be precise about what memory means operationally: the mathematical notion of memory is defined as the net change in the wave between the asymptotic past and the asymptotic future --- i.e., before and after the burst, separated by an \textit{infinite} amount of retarded time.
In this sense, the memory begins building up from the very first moment of the source's evolution, growing imperceptibly slowly at first and ramping up to an eventual ``jump'' at merger time. 
The memory that is actually detectable by gravitational-wave detectors, however, is this characteristic ramp up and jump to produce a net change within a \textit{specific}, finite amount of time --- typically the duration over which most of the gravitational-wave energy is radiated.\footnote{No real detector has sensitivity down to zero frequency, whether due to seismic or thermal noise walls, finite mission duration, or laser-frequency-noise cancellation built into the instrument such as arm-locking or time-delay mechanisms. 
Consequently, a detector is blind to the asymptotic value
the memory departs from or approaches to, and is sensitive only to its time-dependent buildup~\cite{Zosso:2026czc, Inchauspe:2024}; this is precisely the finite-duration ramp that we study here.}
Explicitly, as we shall see later, 
the typical memory wave in LISA has a characteristic rise time of around $70 \frac{GM}{c^3} \sim 300\mathrm{s} \left(\frac{M}{10^6 M_{\odot}}\right)$ (for equal mass binaries) \cite{Favata:2009ii}. 
Even being generous with this timescale by inflating it via the mass ratio, high total mass mergers, redshift, etc., one is fundamentally limited to memory signals with a ramp-up timescale shorter than about 
$10^4-10^5\mathrm{s}$. It would thus seem that, in general, in excess of tens of thousands of supermassive black hole mergers each year are needed for these signals to overlap in the time-domain, or in the most favorable scenario closer to a thousand. On the other hand, according to current expectations LISA would detect only between one to several hundreds of such binaries per year \cite{LISA:2022yao}.

The arrival statistics of these memory events set requirements for two aspects of LISA data analysis: extraction of the memory signal itself, and its treatment within the LISA global fit.
Recent interest in stochastic memory backgrounds~\cite{Zhao:2021zlr, Boybeyi:2024aax} 
motivates a clear characterization of the unresolvable part of the memory spectral energy density within the LISA context, extending existing results for 
continuous-wave backgrounds \cite{Rosado:2011kv}. Unresolved memory signals also serve as a physically motivated test case for the broader problem of residuals from imperfect subtraction of loud sources in the global fit \cite{Rosati:2024lcs, Katz:2024oqg}. 
The difficulties of a global fit and the interest in cosmological populations of supermassive black hole mergers make it interesting to look for more direct ways to extract population characteristics  \cite{Srinivasan:2025etu}. 

In this work, we forecast the stochastic background of gravitational-wave memory in the LISA band, providing initial steps towards handling this aspect of the LISA data stream.
The paper is organized as follows. Section~\ref{sec:singleevents} 
reviews the mathematical foundations of single memory events, reproduces known SNR benchmarks, and extends these to our own forecasts. Section~\ref{sec:population} reviews the current state of the art in SMBH population modeling and introduces the population models used in this work. 
Section~\ref{sec:stochasticbackground} forecasts the stochastic memory background from a LISA observing run, characterizes its statistical nature and expected SNR range, and discusses the implications we find for the LISA mission.

\paragraph{Notation and conventions} For the convenience of the reader, we will introduce all quantities as we go. 
Unless explicitly included, we use $G=c=1$ units throughout this work. 
We will refer to SGWMB as background and stochastic background interchangeably between all three. We will state explicitly the origin of any stochastic background mentioned that does not have to do with memory. 

\section{Individual Event Memory}\label{sec:singleevents}
\label{sec:individual_event_mem}

In this section, we describe the gravitational wave memory strain from an individual binary black-hole merger event, its dependence on source parameters, and its detectability.

\begin{figure}[t]
  \centering
  \includegraphics[width=0.49\textwidth]{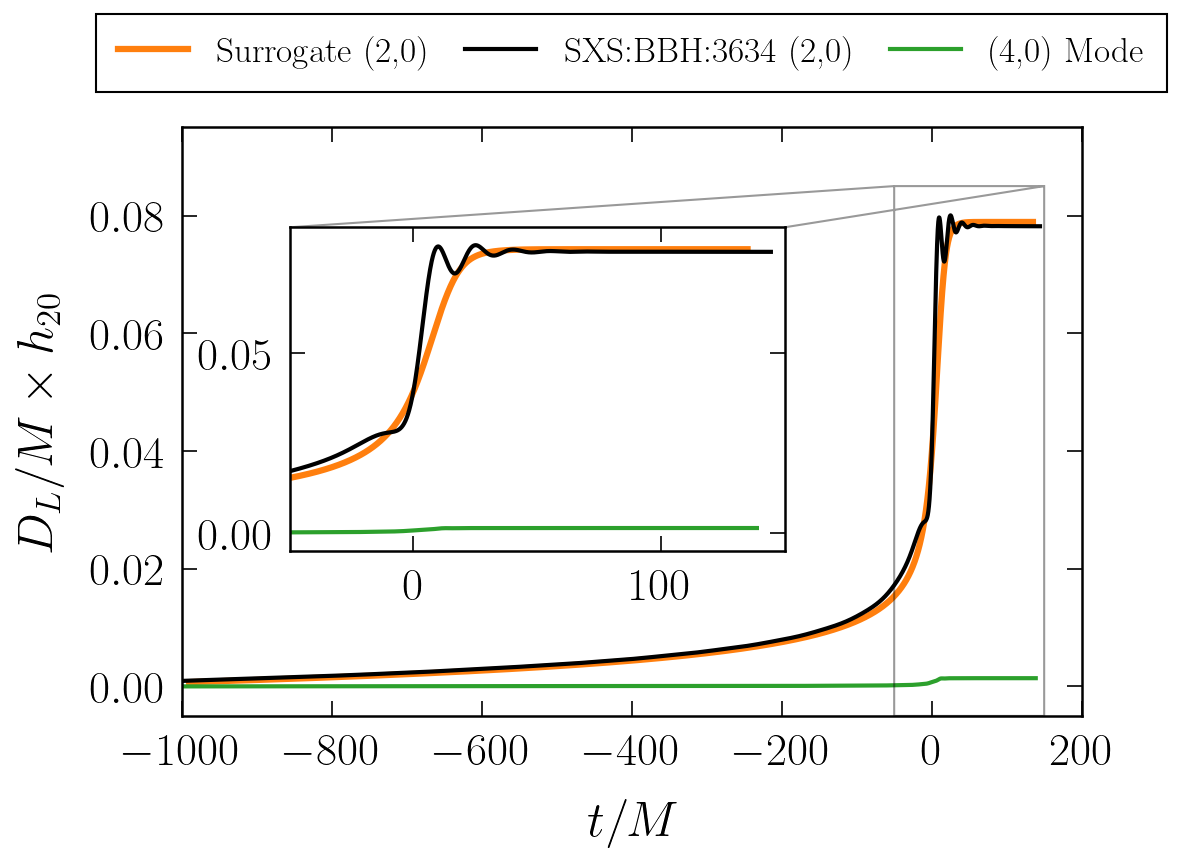}
  \caption{Dimensionless amplitude and time arrays for a BH binary that is non-spinning ($\chi_1, \chi_2 \approx 0$) and with mass ratio $q\approx1$. We present comparisons between both the (2,0) and (4,0) modes as well as between the surrogate model \texttt{NRHybSur3dq8-CCE} and a relevant Numerical Relativity waveform from the SXS collaboration, \texttt{SXS:BBH:3634}.}
  \label{fig:memtimedomain}
\end{figure}

\subsection{Gravitational-wave strain}
\label{sec:2:strain}

At an angular position $(\theta,\phi)$ at future null infinity, it is conventional to organize the two gravitational-wave polarizations, $h_+$ and $h_\times$ into a complex time series of $h_+ - i h_\times$~\cite{RevModPhys.52.29, Kidder:2007rt}.
Using the $-2$ spin-weighted spherical harmonics,
also as a function of retarded time $t$, the complex $h(t,\theta,\phi)$ can be decomposed into a sum  over spin-weighted spherical harmonics:
\begin{align}
h(\theta,\phi,t) =     h_{+} - ih_{\times} = \sum_{l=2}^{\infty}\sum_{m=-l}^{l} {_{-2}}Y_{l,m}(\theta,\phi) h_{l,m}(t),
\end{align}
where the mode amplitudes $h_{lm}(t)$ in turn 
depend on all the parameters of all the sources under consideration. In particular, for a single binary merger this includes {\it intrinsic parameters} such as total mass $M$, mass ratio $q = m_1/m_2$ (with $m_1> m_2$), the black hole spins, and orbital parameters at a reference time, as well as {\it extrinsic parameters} such as wave emission direction relative to the source. 
For the purposes of our analysis so far, we will simply consider quasi-circular orbits and aligned spins, therefore restricting ourselves to four intrinsic parameters: the total mass $M$, mass ratio $q$, as well as the the dimensionless $\chi_{1}$ and $\chi_{2}$. We further constrain ourselves in these parameters following the analysis in Sec.~\ref{sec:source_param_dep}. 

The predominant quadrupolar oscillatory modes are in $(l,m)=(2,\pm2)$ in the spin-weighted spherical harmonic basis, with 
\begin{equation}
\left.
\begin{aligned}
_{-2}Y_{2,2}(\theta,\phi) &= \sqrt{\frac{5}{4\pi}} \cos^4 \left(\frac{\theta}{2}\right)e^{2i\phi}\, , \\
_{-2}Y_{2,-2}(\theta,\phi) &= \sqrt{\frac{5}{4\pi}} \sin^4 \left(\frac{\theta}{2}\right)e^{-2i\phi}\, .
\end{aligned}
\right\}
\end{equation}
which are strongest when viewed face on, with $\theta=0$ or $\pi$. 
However, in the post-Newtonian approximation, the gravitational-wave memory effect is dominated by the $(\ell, m) = (2, 0)$ spin-weighted spherical harmonic mode~\cite{Christodoulou:1991, Blanchet:1992, Favata:2009ii}. This post-Newtonian prediction was subsequently confirmed in full numerical relativity: computations converting NR binary black hole waveforms into memory contributions~\cite{Pollney:2010, Talbot:2018sgr} and, more recently, direct extraction via Cauchy-Characteristic Evolution~\cite{Mitman:2020, Mitman:2021, Mitman:2024} consistently find that the $(2,0)$ mode accounts for the overwhelming majority of the memory signal, at least in the aligned-spin configurations considered here. Noting that 
\begin{equation}
\label{profileY20}
_{-2}Y_{2,0}(\theta,\phi) = \sqrt{\frac{15}{32\pi}} \sin^2{\theta}  \, .
\end{equation}
which is real-valued, we will only have (a real-valued) $h_+$ contribution, given by 
\begin{align}
    h_{+}(t) \approx {_{-2}Y_{2,0}(\theta,\phi)} h_{20}(t)
\end{align}
where we have omitted the $(\theta, \phi)$ dependence in $h_+$ for visual clarity.
From Eq.~\eqref{profileY20}, it follows that the memory wave from a spin-aligned binary is the strongest when viewed edge on.

The memory as measured in the detector at merger time can then be written as 
\begin{equation}
    h_{\rm det}(t) = F_{+}(\theta, \phi, \psi) h_{+}(t),
\end{equation}
where the antenna pattern is given as \cite{Zhao:2021zlr, Maggiore:2007ulw}
\begin{multline}
    F_{+}(\theta, \phi, \psi) = -\frac{1}{2}(1 + \cos^2\theta)\cos 2\phi \cos 2\psi \\ -\cos\theta \sin 2\phi \sin 2\psi.
\end{multline}
with its sky averaged polarization \cite{amaroseoane2017laserinterferometerspaceantenna}: 
\begin{align}
    \langle F_{+}^2 \rangle = \frac{1}{4\pi^2} \int_0^{2\pi} d\phi \int_0^{\pi} \sin\theta \, d\theta \int_0^{\pi} d\psi \, F_{+}^2 = \frac{1}{5}.
\end{align}
The functional form of $F_{+}$ is generic to any two-arm strain-projecting
detector. Ground-based interferometers and LISA (treated as an effective Michelson
detector via time-delay interferometry) share this pattern up to an overall geometric rescaling from a $90^{\circ}$ to a $60^{\circ}$ arm opening angle
\cite{Zhao:2021zlr, Maggiore:2007ulw, Cutler:1997ta, Cornish:2002rt}.
Pulsar timing arrays, by contrast, measure a one-way pulsar--Earth redshift rather
than a two-arm strain, giving each pulsar a single-detector response function 
with a
qualitatively different angular dependence. Subsequently, quadrupolar sky sensitivity is instead recovered only after correlating timing residuals between pairs of pulsars, yielding
the Hellings--Downs curve \cite{Hellings:1983fr}.

In this paper, 
for single-event sources, we will use the surrogate models \texttt{NRHybSur3dq8-CCE} from \cite{Yoo:2023spi}, which in particular include the memory component, to approximate the gravitational wave signals. These surrogate models are, at the time of writing, the most up-to-date model and incorporate numerical relativity simulations with $1 \leq q \leq 8$ and $-0.8 \leq \chi_{1z}, \chi_{2z} \leq 0.8$, which capture GW memory effects via Cauchy Characteristic Extraction (CCE)~\cite{Bishop:1996xx}. They are made publicly available through the Python package \texttt{GWSurrogate} \cite{Field:2025isp}. In addition, these results can be verified directly against numerical relativity waveforms the SXS collaboration catalog \cite{Scheel:2025jct}, as made available through the \texttt{sxs} Python package. In Fig.~\ref{fig:memtimedomain} (bottom), we compare the surrogate model we use and the numerically computed \texttt{SXS:BBH:3634} waveform, which describes an equal mass system with nonspinning progenitors, both in the (2,0) mode and as dimensionless quantities in the time-domain. As we can see, the surrogate model suffices to be used as it is in good agreement with the NR simulation.

In Fig.~\ref{fig:memtimedomain}, we illustrate with the modes available in the surrogate models the memory ``effect'' $\Delta h_{lm +/\times}^{\rm mem}$, 
which we define to be the difference between the strain at times $+100M$ and $-5000M$ where the origin is fixed at peak strain.
As is clear from Fig.~\ref{fig:memtimedomain} and Refs.~\cite{Boybeyi:2024aax, Favata:2008ti, Zhao:2021zlr}, the gravitational wave memory is almost entirely in the (2,0) mode of the $h_{+}$ polarization. Therefore, for the remainder of this paper we will approximate and refer to memory jump as 
\begin{equation}
    \Delta h^{\rm mem} \sim \Delta h_+ \approx {_{-2}Y_{2,0}} \Delta h_{20}.
\end{equation}

\begin{figure*}[t]
  \centering
\includegraphics[width=\textwidth]{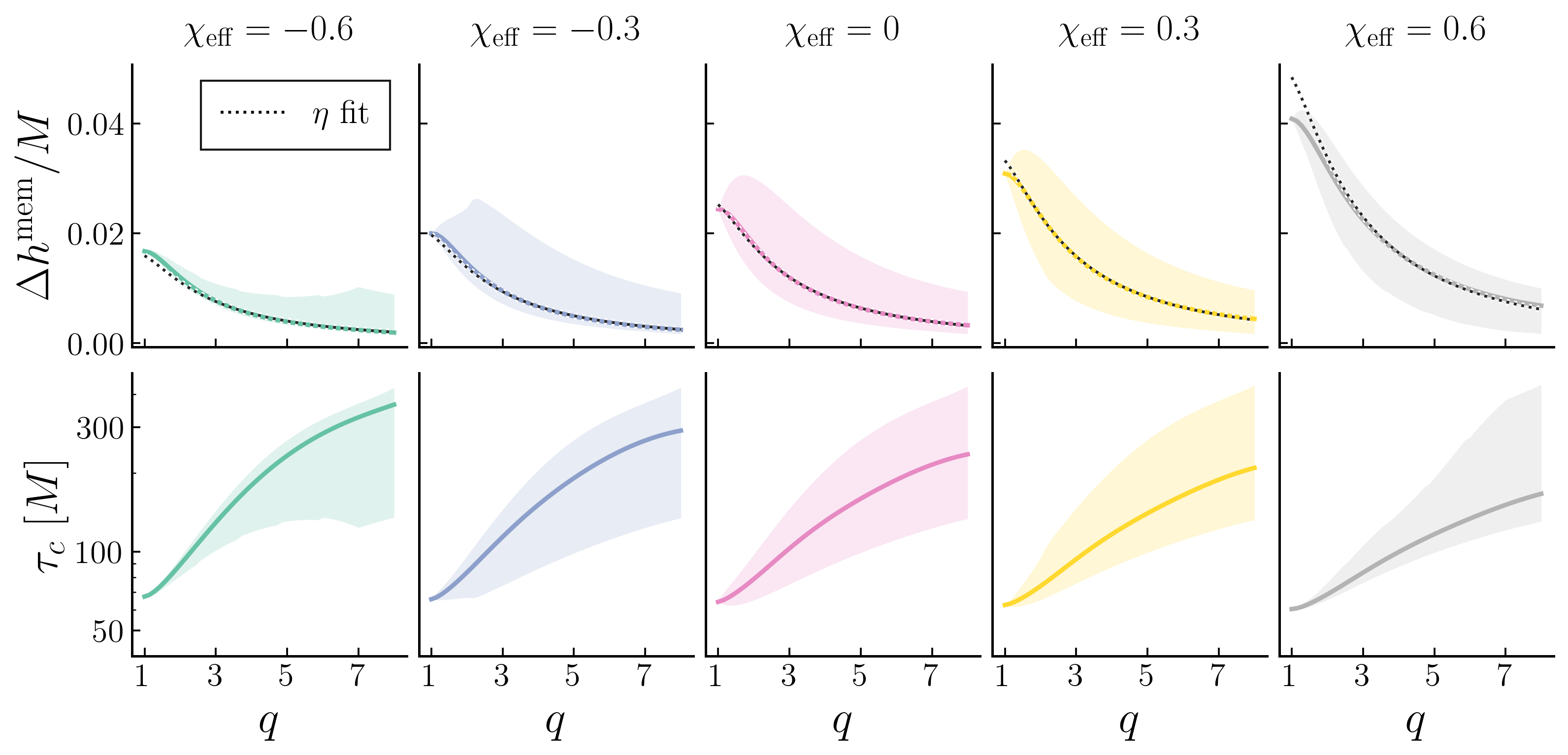}
\caption{Gravitational wave memory parameters as a function of mass ratio $q \in [1, 8]$ for five values of the effective spin parameter $\chi_\mathrm{eff}$, evaluated at inclination $\iota = \pi/3$. Solid lines show the median and shaded bands span the range of $(\chi_1, \chi_2)$ combinations that realize each fixed $\chi_\mathrm{eff}$. \textit{Top:} For all spin values, the memory strain jump $\Delta h^{\rm mem}$ 
  trends with $\eta$. 
  We highlight this with dashed guide curves showing pure symmetric-mass-ratio scaling, $\propto \eta = q/(1+q)^2$, normalized to the median value at $q=1$ for each $\chi_\mathrm{eff}$. The departure of the medians from these guides indicates the modest deviation from pure $\eta$ scaling over the plotted $q$ range. 
  Higher $\chi_\mathrm{eff}$ produces larger memory amplitudes, with the bands at fixed $\chi_\mathrm{eff}$ reflecting the residual dependence on the individual spin components $\chi_1$ and $\chi_2$. \textit{Bottom:} In contrast to $\Delta h^{\rm mem}$, the characteristic rise time $\tau_c$ increases monotonically with $q$ and is larger for more negative $\chi_\mathrm{eff}$. The bands widen substantially at large $q$, indicating that the individual spin decomposition becomes increasingly important for the rise time at asymmetric mass ratios. We note that the \textit{specific} values of $\tau_c$ will vary by what one defines as the exact window the jump of the memory is constrained to. 
  }
  \label{fig:memorycharacteristics}
\end{figure*}

Because the vacuum field equations are scale-invariant when $G=c=1$, the waveform of a binary with fixed mass ratio $q$ and aligned spin $\chi_z$ depends on time only through $t/M$: rescaling the total mass $M$ rescales
the system's length and time scales together, leaving the dimensionless waveform shape unchanged. The parameters $q$ and $\chi_z$ are themselves dimensionless and are unaffected by this rescaling, so they enter as fixed labels selecting which member of this universal family applies, i.e.
$H_{20}(t/M;\,q,\chi_z)$.
As the wave propagates over cosmological distances, its time argument is stretched by the redshift, $t/M \to t/M_z$ with $M_z \equiv M(1+z)$, and
its amplitude falls off as $1/D_L(z)$. 
In practice we employ these scalings on numerically computed waveforms from the surrogate model \texttt{NRHybSur3dq8-CCE} \cite{Field:2025isp, Talbot:2018sgr} to uncover our presented results.

Beyond its role in amplitude calibration, the luminosity distance is a fundamental cosmological observable that relates source distances to the expansion history of the Universe.
It can be calculated in terms of the redshift as $D_L = (1+z) d(z)$, where $d(z)$ is the co-moving distance
\begin{align}
  d(z) = \cfrac{c}{H_0} \int_{0}^{z} \cfrac{dz'}{\sqrt{\Omega_{\Lambda} + \Omega_M (1+z')^3}} \, .
\end{align}
Here, $H_0$, $\Omega_{\Lambda}$, and $\Omega_{M}$ are cosmological parameters for which we used the $\Lambda$CDM informed by Planck data. In practice, we compute the luminosity distance using the Planck18 cosmology from the Python package \texttt{astropy} \cite{astropy:2013}.

\subsection{Dependence on source parameters}
\label{sec:source_param_dep}
The memory gravitational wave from a binary black-hole merger event can be characterized by the magnitude of the jump $\Delta h^{\rm mem}$ and the corresponding time $\tau_c$ it takes to rise. 
In light of Fig.~\ref{fig:memtimedomain} we restrict our definition of the memory jump $\Delta h^{\rm mem}$ (see Sec.~\ref{sec:2:strain}) to only consist of strain extracted from the dominant (2,0) mode of the merger.
Likewise, we define $\tau_c$ as the time interval in which the memory amplitude dramatically increases, or the full-width half-maximum of the peak energy flux $dE/dt$. 
We explore both of their dependences on source parameters, namely $(M,q,\chi_{1,2})$, which we draw scaling information from Post-Newtonian models and extract accurate values from the \texttt{NRHybSur3dq8-CCE} surrogate model.  Our results are summarized in Fig.~\ref{fig:memorycharacteristics}, 
where we plot the strain jump and rise time as a function of $q$ and $\chi_\text{eff}$. 

Post Newtonian studies provide an analytic scaling to important parameters, such as $h_c$ and $\tau_c$, founded on common source parameters, such as $q$, $\chi_{\rm eff}$, $D_L$, $M$, etc. For example, the plus polarization can be written from the radiated energy using the leading order multiple moment approximation as
(see~\cite{Favata:2009ii}, Eqs.~(5-7) and \cite{Pollney:2010}, Eqs.~(9 \& 10)) \footnote{In Ref.~\cite{Favata:2009ii}, the memory amplitude is referred to simply as $h^{\rm mem}$ instead of $h^{\rm PN, mem}$. Here we specifically denote with ${\rm PN}$ to show we are considering the author's equations of Post-Newtonian origin in comparison to our analysis in numerical relativity. 
Furthermore, we choose this to avoid confusion in named ``$h$'' terms with the rest of our text.}
\begin{equation}
h_{+}^{\rm PN, mem}(t) \approx \frac{\eta M h^{\rm PN, mem}(t)}{384\pi R} \sin^2\theta\left(17 + \cos^2\theta\right),
\label{eq:pn1}
\end{equation}
with $\eta = m_1 m_2/M^2=q/(1+q)^2$ the symmetric mass ratio, and
\begin{equation}
h^{\rm PN, mem}(t) \sim \frac{1}{\eta M} \int^t \cfrac{dE_{\rm rad}}{dt'}\, dt'
\label{eq:pn2}
\end{equation}
where the radiated energy itself scales (roughly, and omitting the time dependence found in Eq. 6 of Ref.~\cite{Favata:2009ii}) as
\begin{equation}
    E_{\rm rad} \propto \sqrt{\cfrac{8 \pi}{5}}\eta M,
    \label{eq:pn3}
\end{equation}
along with factors of orbital frequency and separation~\cite{Favata:2009ii}.
The (2,0) mode amplitude can likewise be approximated as a power series in symmetric spin $\chi_{\text{eff}} \equiv (m_1\chi_{1z} + m_2\chi_{2z})/M$ as~\cite{Pollney:2010, Liu:2021zys, Zhao:2021zlr}
\begin{multline}
    \frac{D_L}{M} \Delta h_{\rm 20} \approx 0.0969 + 0.0562\chi_{\text{eff}} + 0.0340\chi^{2}_{\text{eff}} + \\ 0.0296\chi^{3}_{\text{eff}} + 0.0206\chi^{4}_{\text{eff}}.
\end{multline}
Overall, post-Newtonian calculations for the jump magnitude $\Delta h^{\rm mem}$ indicate it approximately scales as   
\begin{equation}
    \Delta h^{\rm PN, mem} \sim \eta M f(\chi_{1},\chi_2)\,.
\end{equation}
From the surrogate model \texttt{NRHybSur3dq8-CCE}, for each fixed spin configuration, we indeed confirm the decrease of $\Delta h^{\rm mem}$ with increasing  $q$ (with $q\ge 1$).  In particular,  this leads to a suppression by factor of $\sim 10$ when $q$ reaches 8. 
We also find that $\Delta h^{\rm mem}$ can double if both black holes are rapidly spinning --- and aligned with the orbital angular momentum.  Both trends confirm  previous studies in Refs.~\cite{Favata:2009ii, Pollney:2010, Talbot:2018sgr, Liu:2021zys, Yoo:2023spi, Elhashash:2024thm}.
Our results are shown in the upper panel of Fig.~\ref{fig:memorycharacteristics}.

Since the memory strain is proportional to a time integral of the radiated
energy flux (Eq.~\eqref{eq:pn2}), its
instantaneous growth rate tracks the instantaneous GW luminosity
\cite{Christodoulou:1991, Blanchet:1992, Favata:2009ii}.
The rise time $\tau_c$ is therefore set by the duration over which most
of the binary's energy is emitted.
These studies predict that $\tau_{\rm c} \propto M$, governed primarily by the energy emitted near merger for equal-mass, non-spinning systems. 
As we see in the bottom panel of Fig.~\ref{fig:memorycharacteristics}, for a fixed total mass $M$, $\tau_{\rm c}$ can increase by a factor from $2$ to $5$ as mass ratio $q$ increases towards 8.  Separately, having aligned spins also tend to increase $\tau_c$, especially in binaries with asymmetric masses. 

While the longer duration and distinct frequency contribution might suggest high-$q$ events are interesting for an unresolvable background, their overall contributions to memory signal are ultimately subdominant. The power of the memory background scales with both the astrophysical merger rate and the square of the memory jump amplitude. Because the jump amplitude drops so significantly for $q > 1$, this effect overwhelmingly suppresses the contribution from high mass-ratio binaries. Consequently from Fig.~\ref{fig:memorycharacteristics} (see also \cite{Islam:2021old,Elhashash:2024thm,Cunningham:2024dog}), 
we expect the stochastic GW memory background to be dominated by the stronger and more frequent mergers of comparable-mass binaries ($q \approx 1$). Similarly, we restrict our analysis to non-spinning binaries. Spin effects primarily
influence the inspiral dynamics and higher-order waveform harmonics,
whereas the memory amplitude is dominated by the total energy radiated
during merger. In Fig.~\ref{fig:memorycharacteristics}, certain choices of aligned spin configurations can modestly enhance or suppress $\Delta h^{\rm mem}$, but this effect is, again, ultimately subdominant and unimpactful when considering large numbers of events, as one would do for modeling a stochastic background of memory. We therefore restrict the remainder
of our analysis to non-spinning, equal-mass binary mergers. 

\subsection{Frequency-Domain Signal and Detectability of Individual Events}

Assuming the detector to have stationary Gaussian noise and the matched-filtering technique is used, the SNR achievable for a single GW event with frequency-domain characteristic strain $h_c^{\rm mem}$ is given by~\cite{Favata:2009ii}: 
\begin{align}
    \text{SNR} = \sqrt{\int_0^{\infty} \cfrac{|h_c^{\rm mem}|^2}{|h_n|^2} \cfrac{df}{f}}
\end{align}
where $h_n$ is the sky averaged rms noise amplitude given as $h_n(f) = \sqrt{\frac{20}{3}fS_n}$ for the noise spectral density $S_n$. The characteristic strain itself $h_c$ is given in terms of the Fourier transform of the plus-polarization strain
$\tilde{h}_+$ by
\begin{align}
    h_c^{\rm mem}(f) = 2f(1+z)\times \langle |\tilde{h}_+^{\rm mem}(f(1+z))|^2 \rangle^{1/2}_{r\rightarrow D_L(z)} \, ,
\end{align}
where the angled brackets $\langle \cdot \rangle$ denote an average over sky and polarization angles. 

\begin{figure}[t]
  \centering
  \includegraphics[width=0.49\textwidth]{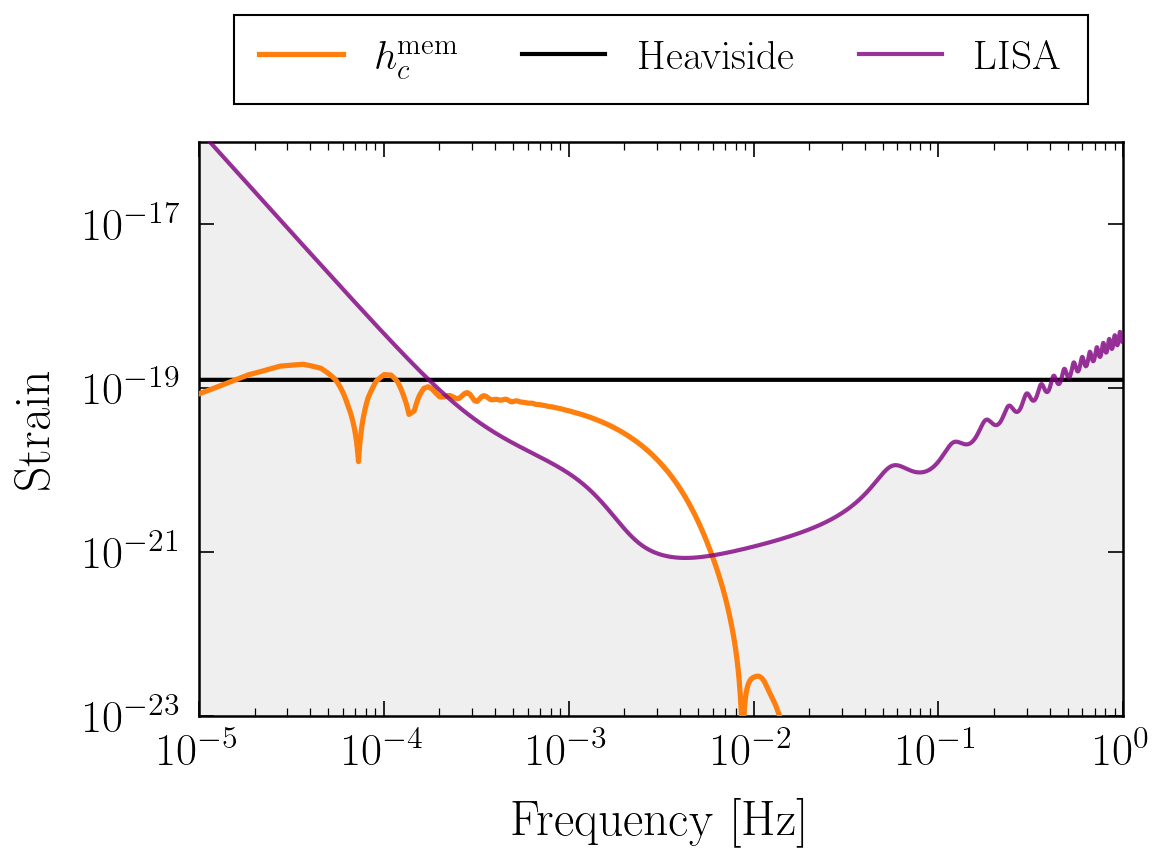}
  \caption{The GW memory waveform in the frequency domain. In orange is the corresponding characteristic strain and in black is a reference using a Heaviside function approximation to the memory waveform. They are both plotted against the LISA sensitivity curve in purple. The parameters of the merger are $z=1.0$, $M=10^6 M_\odot$, $\theta=\pi/3$.}
  \label{fig:memfreqdomain}
\end{figure}

As we move into the frequency domain, a time-domain strain of $h^{\rm mem}(t)$ leads to a $\sim 1/f$ power-law dependence at low frequencies, while the finite rise time $\tau_c$ leads to a cutoff of $f_{\rm c} \approx 1/\tau_{\rm c}$, above which $|h(f)|$ rapidly falls off. For this reason, given the same $h^{\rm mem}(t)$, faster-rising signals will contribute more toward gravitational-wave energy content by extending into higher frequencies.   
We note that previous treatments of GW memory in the LISA band \cite{Zhao:2021zlr, Zhang:2025rux, Boybeyi:2024aax, Pollney:2010, Liu:2021zys} have modeled the time-domain memory signal as a Heaviside step function, which corresponds to an instantaneous rise time ($\tau_{\rm c} \to 0$) and therefore artificially extends the $1/f$ power law to arbitrarily high frequencies with no cutoff. 
In Fig.~\ref{fig:memfreqdomain}, we illustrate the Fourier transform of characteristic memory signal strain with the parameters: angle of inclination $\theta = \pi/2$, redshift $z = 1.0$, and total mass $M=10^6 M_\odot$ together with a step-function approximation and the LISA sensitivity. 
Clearly, the Heaviside approximation to the memory significantly overestimates the characteristic strain at frequencies above $f_{\rm c}$ when compared to NR waveforms, and we caution that conclusions drawn from such an approximation about SNR or detectability may be overly optimistic. 

While using NR waveforms are crucial for precise forecasts of the memory SNR, input of that alone is not enough for accurate estimates. 
In practice, we compute the LISA noise power spectral density $S_n$ (and
correspondingly $h_n$) using \texttt{legwork} \cite{Wagg:2021sgn}, which
accounts for the detailed structure of the LISA detector response and
also provides SNR calculations for individual events. Furthermore, before Fourier
transforming the time-domain signal, we extend it at its final value and
apply a Planck-taper window \cite{planck:2010tapering}, which
suppresses ringing while preserving the waveform's finer features.

\begin{figure}[t]
  \centering
  \includegraphics[width=0.49\textwidth]{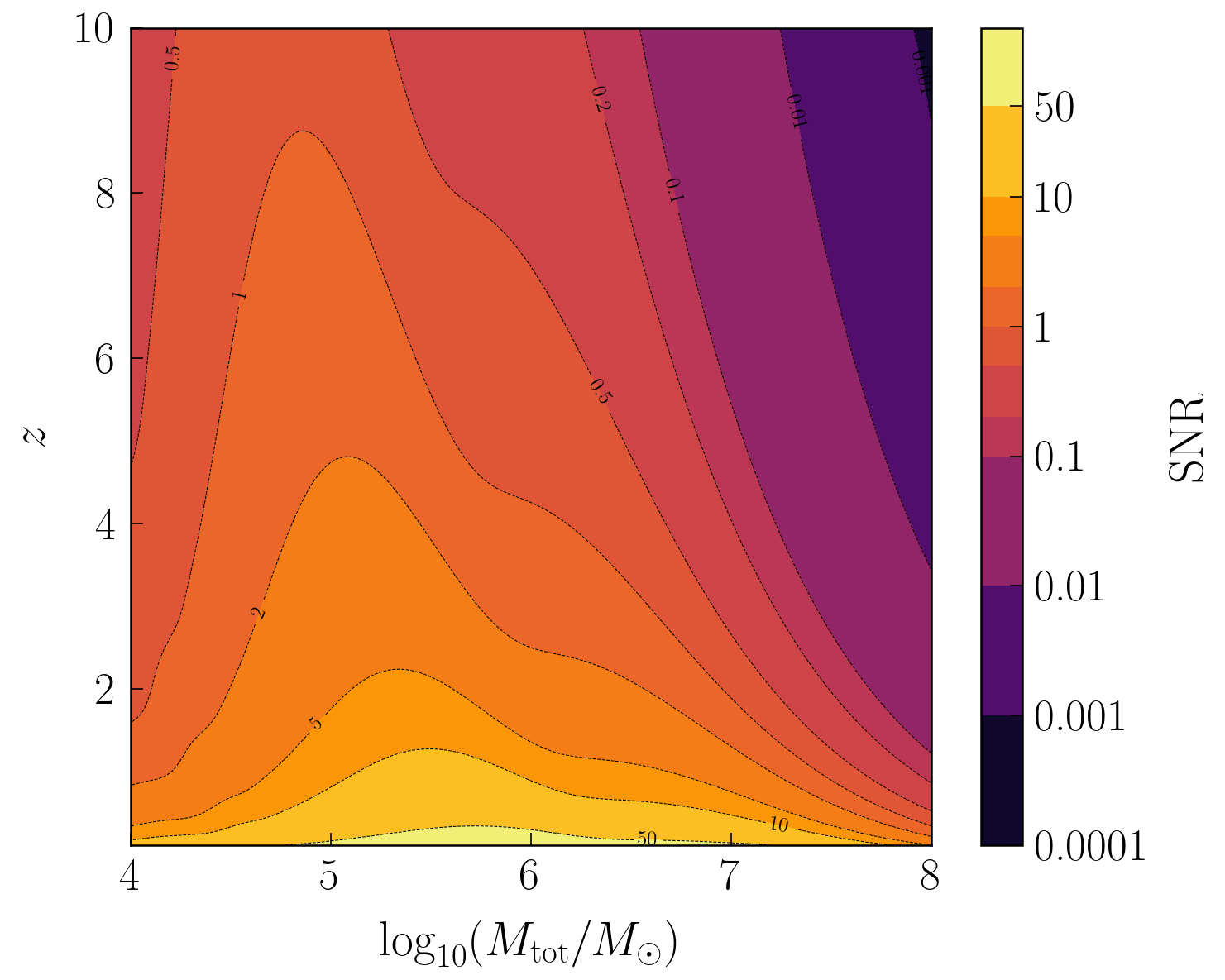}
  \caption{Contour plot of the parameter space of redshift and total mass, with contour lines denoting the memory SNR. All mergers have inclination $\pi/3$.
  We find the magnitude and shape are consistent with results with \cite{Inchauspe:2024}.}
  \label{fig:ind_mem_heatmap}
\end{figure}

In Fig.~\ref{fig:ind_mem_heatmap}, we forecast the detectability of the gravitational-wave memory as individual SMBH mergers within the mass range of $M=[10^4-10^{8}]M_\odot$ and redshift range of $z=[0.1-10]$.
We find that the individual memory SNR in LISA can be expected to be of order $\mathcal{O}(0.1-1)$ in most cases, with an approximate peak SNR of 67\footnote{This maximum SNR will vary according to smoothing and windowing techniques performed on the memory waveform. We do not present this as a final, global maximum of memory SNR and instead as a reference value to the true maximum, which will be of order $\mathcal{O}(100)$.}
at redshift $z = 0.1$. 
To place these results in their proper astrophysical context, it is useful to compare the memory SNR to the SNR of the full, oscillatory waveform from the same sources. 
While the memory signal shown to be individually quite weak in most circumstances by Fig.~\ref{fig:ind_mem_heatmap}, the full Inspiral-Merger-Ringdown (IMR) waveforms of the corresponding SMBH events are expected to be resolved at SNRs of order $\mathcal{O}(100-1000)$ out to high redshift by the LISA detector  (see the leftmost plot of Fig. 6 in Ref.~\cite{Inchauspe:2024}, as well as~\cite{Spadaro:2026evb, Piarulli:2025rvr, Toubiana:2023cwr, Katz:2018dgn}).
As a result, only the memory from the loudest and most nearby events will be detectable as an individual burst. This finding is fully consistent with recent, more detailed analyses on the subject (see the middle plot of Fig. 6 in Ref.~\cite{Inchauspe:2024}). 
This gives us our proper motivation to search for memory in the form of a stochastic background. Since the vast majority of memory events will be individually sub-threshold, their accumulated signals could in principle contribute a detectable foreground within LISA --- though whether this background is truly continuous or merely a sparse sequence of discrete bursts depends critically on the underlying SMBH population, and is the central question we address in the remainder of this paper.

\section{Population Models}\label{sec:population} 
A number of
factors 
govern SMBH population models and directly shape both the redshift evolution and the mass, mass–ratio, spin, and eccentricity distributions of binaries. They
include the seeding channel (light Population III remnants vs. heavy direct–collapse seeds)~\cite{Begelman:2006db, Rees:2007nc, Inayoshi:2019fun} and the accretion mode (coherent/efficient vs.\ chaotic/bursty episodes)~\cite{King:2006uu}, which affects spins and growth duty cycles as well as pairing and hardening physics (dynamical friction, stellar loss–cone scattering, circumbinary–disc torques)~\cite{Berti:2008af, Volonteri:2007tu, Begelman:1980vb}. In turn, these set merger delays~\cite{Kelley:2017lek}, eccentricity at band entrance~\cite{Wang__2025, LISA:2022yao}, and the occupation fraction and scaling–relation scatter in low–mass hosts, which impact the low–mass cutoff and the mass–ratio spectrum~\cite{Volonteri_2009, Miller:2014vta}. 

Gravitational-wave observations offer a probe of this population that spans a wide range of frequencies and is largely independent of AGN activity or luminosity. At $\sim 10^{-9}\,\mathrm{Hz}$, pulsar--timing arrays (PTAs) now report Hellings--Downs--correlated evidence for a stochastic gravitational--wave background consistent with an ensemble of SMBH binaries \cite{NANOGrav:2023gor}.
PTA measurements thereby provide an external constraint on the overall background amplitude and any environmental deviations from vacuum inspiral, while LISA-band inference constrains the mass spectrum, redshift distribution, and merger rate of the underlying binary population. 
Together, these two observational windows offer complementary and mutually reinforcing handles on the SMBH merger history.
Here we adopt two population models that bracket the current astrophysical uncertainty: a flexible parametric model whose hyperparameters are constrained by PTA observations, and a physically-motivated model anchored to the dark-matter halo mass function and empirical black-hole--halo scaling relations.
It should be noted that the uncertainty within Model 1, as well as other SMBH population models such as \cite{Langen:2024ygz, Chen:2018znx, Barausse:2020kjy}, 
is still quite large. This uncertainty arises from our current limited understanding of SMBH populations derived from cosmological simulations, to which we are optimistic that results from LISA will provide much-needed constraints. Such constraints would have significant implications for not only our understanding of gravitational waves but also galactic evolution \cite{sun2025mmrmbhrelationevolution, Fan:2024nnp, zou2024cosmicevolutionsupermassiveblack}.

\begin{figure*}[t]
  \centering
  \includegraphics[width=0.99\textwidth]{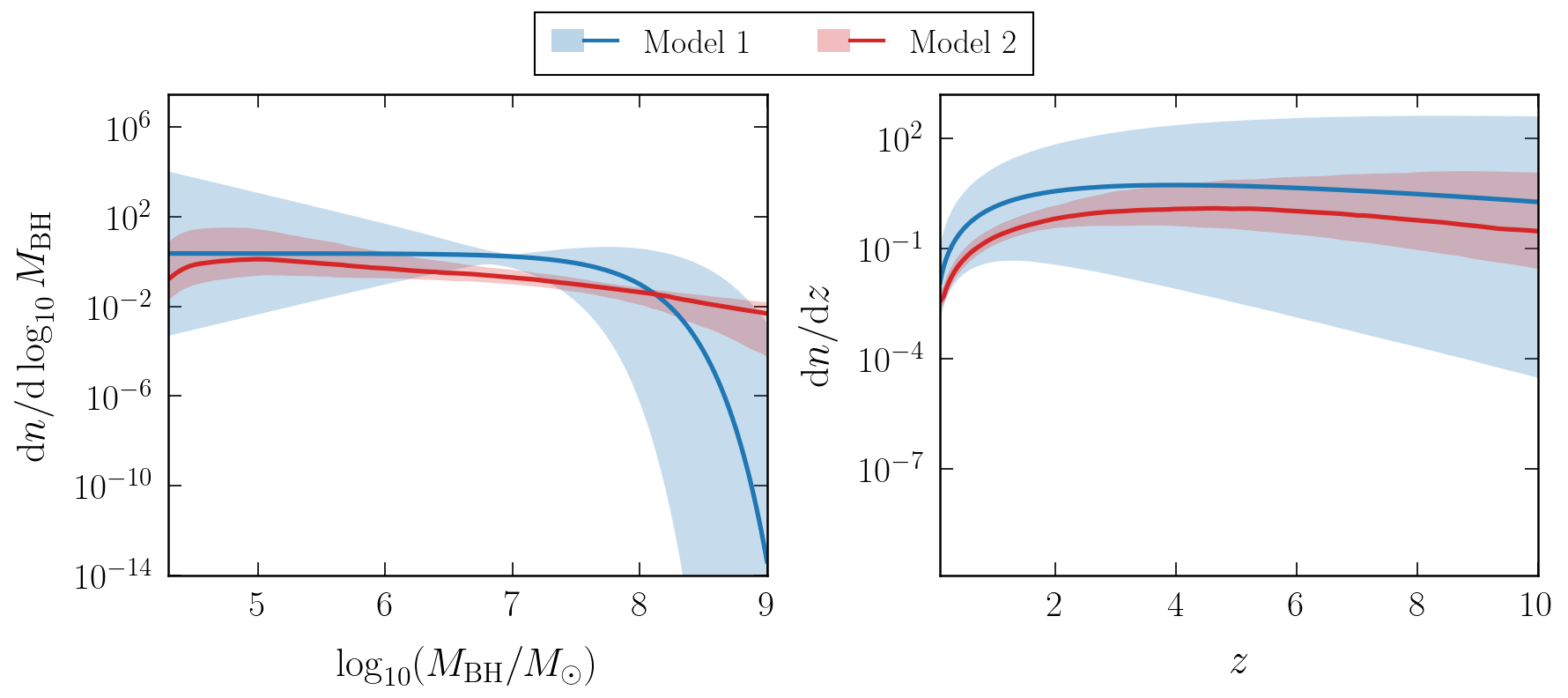}
  \caption{SMBH merger rate distributions for Model 1 (blue) and Model 2 (red)
across the Monte Carlo ensemble, shown as median curves with $1\sigma$
shaded bands in parameter space for the respective model. \textit{Left:} differential merger rate per unit
$\log_{10}$ mass, $\mathrm{d}n/\mathrm{d}\log_{10} M_\mathrm{BH}$.
\textit{Right:} differential merger rate per unit redshift,
$\mathrm{d}n/\mathrm{d}z$. Model 1's broader uncertainty at high mass
and high $z$ reflects its wide prior on parametric hyperparameters, and
is the primary driver of the broader SNR and overlap distributions as opposed to Model 2 shown in
our later analysis.}
  \label{fig:model_comparison}
\end{figure*}

\subsection{Model 1} 
As a first model, we will adopt the flexible analytic model outlined in Refs.~\cite{middleton2015astrophysical, middleton2018no, steinle2023implications}.
In this approach, the differential merger rate density with respect to comoving volume, redshift $z$, and mass $M$ is described by a single function, $\mathcal{R}(z, M)$.

For 
simplicity, we adopt a factorized form of this model, following the implementation presented in Refs.~\cite{Boybeyi:2024aax, Zhang:2025rux, middleton2015astrophysical}, which ignores the evolution of mass distribution over through redshift. $g_{\text{SMBH}}(M)$ is defined to be a normalized probability density function, such that $\int g_{\text{SMBH}}(M) \mathrm{d}\log M = 1$, and  $R_{\text{SMBH}}(z)$ represents the total comoving merger rate density at a given redshift; hence in this model $\mathcal{R}=R_\text{SMBH}g_\text{SMBH}$. We note that while
Refs.~\cite{middleton2015astrophysical} often formulate the mass distribution in terms of the chirp mass ($\mathcal{M}_c$), which is more directly constrained by gravitational wave observations, the model we adopt here is expressed in terms of the total mass $M = M_1 + M_2$. The redshift-dependent rate $R_{\rm SMBH}(z)$ and the mass distribution $g_{\text{SMBH}}(M)$ are subsequently parameterized as \cite{Boybeyi:2024aax, middleton2015astrophysical}:
\begin{align}
    R_{\text{SMBH}}(z) &= \dot{n}_0 (1 + z)^{\beta_z} e^{-z / z_c} \\
    g_{\text{SMBH}}(M) &= \mathcal{N} \frac{1}{M} \left( \frac{M}{10^7 M_\odot} \right)^{-\alpha_M} e^{-M / M_*}
\end{align}
Here the prefactor $\dot{n}_0$ is the local merger rate density (in $\text{Gpc}^{-3} \text{yr}^{-1}$) at $z=0$; the profile of $R_{\rm SMBH}(z)$'s redshift evolution is then governed by a power-law index $\beta_z$ which describes an initial increase at low $z$, and an exponential cutoff with a characteristic redshift $z_c$. Similarly, the mass distribution is described by a power law with a negative slope $\alpha_M$ and an exponential cutoff above a characteristic mass $M_*$. The term $\mathcal{N}$ is a normalization constant chosen to ensure that $g_{\text{SMBH}}(M)$ integrates to 1. In Fig.~\ref{fig:model_comparison}, we plot (in blue) the number density
$dn/dz$ and $dn/d\log M$ 
derived from $R(z)$ and $g(M)$ \cite{middleton2015astrophysical} generated from the mean values of these parameters as well as the $1\sigma$ confidence intervals for these distributions. These intervals were derived from the posterior distributions obtained via the Bayesian analysis performed in Ref.~\cite{Boybeyi:2024aax} using recent pulsar timing array data.
We find the uncertainty in Model 1 by performing Monte-Carlo sampling over the parameter space outlined in Refs~\cite{middleton2015astrophysical, middleton2018no, steinle2023implications}.

\subsection{Model 2} 
We next consider the population model of SMBHs presented in Ref.~\cite{Langen:2024ygz}. In contrast to Model 1, this model is heavily informed by the connection between SMBHs and dark matter (DM) halos \cite{Padilla:2020sjy, Sabra_2015, Booth_2010, Robertson_2006, Delvecchio_2019}. 

The scaling relation between MBH mass $M_\mathrm{bh}$ and host halo mass $M_\mathrm{h}$ is modeled as a broken power law with slopes $\gamma$ and $\gamma'$ above and below a transition halo mass $M_{\mathrm{h},t}$, an overall normalization $\epsilon$, and redshift-dependent corrections involving the matter density parameter $\Omega_\mathrm{m}^z$ and the collapse overdensity $\Delta_c$. The accompanying occupation fraction is assumed to be constant through redshift and is also dependent on the parameters addressed above. We refer the reader to Ref.~\cite{Langen:2024ygz} for the explicit functional forms of these relations and the fitted values of all parameters.

Using the occupation fraction and the halo--MBH mass scaling relation, we write the total merger density with respect to logarithmic mass, redshift, and mass ratio. Following Ref.~\cite{Langen:2024ygz}, we have
\begin{align} \label{dn_informed}
\left( \frac{\mathrm{d}n_{\mathrm{bh}}}{\mathrm{d} \log M_{\mathrm{bh}} \, \mathrm{d}z \, \mathrm{d}q} \right)
&= f_{\mathrm{bh}}(M_{\mathrm{h},1}, z) \, f_{\mathrm{bh}}(M_{\mathrm{h},2}, z) \notag \\
&\quad \times A_1 \left( \frac{M_{\mathrm{bh}}}{10^{12} M_\odot \times K(z, \gamma, \epsilon)} \right)^{\frac{3\alpha}{\gamma}} \notag \\
&\quad \times q^{\frac{3}{\gamma} - 1 + \frac{3\beta}{\gamma}} (1 + z)^{\eta} \times 
\exp\left[ \left( \frac{q}{\bar{q}} \right)^{\frac{3\gamma_1}{\gamma}} \right] \notag \\
&\quad \times \frac{\mathrm{d}n_{\mathrm{h}}}{\mathrm{d} \log M_{\mathrm{h}}}(z),
\end{align}
where $M_{\mathrm{h},1}$ and $M_{\mathrm{h},2}$ are the halo masses corresponding to the primary and secondary black holes (obtained via the inverse of the scaling relation), $q$ is the binary mass ratio, and $\mathrm{d}n_\mathrm{h}/\mathrm{d}\log M_\mathrm{h}$ is the halo mass function. The parameters $A_1$, $\alpha$, $\beta$, $\eta$, $\bar{q}$, and $\gamma_1$ characterize the merger rate amplitude, mass-function slope, mass-ratio distribution, redshift evolution, and mass-ratio cutoff respectively, and are given in Ref.~\cite{Langen:2024ygz}. Moreover, the redshift dependent shorthand $K(z, \gamma, \epsilon)$ encapsulates the normalization and redshift dependence of the halo--MBH mass scaling relation, which we directly source from Ref.~\cite{Langen:2024ygz}. 
This expression assumes that the merger rate of MBHs can be related to the merger rate of their host halos, and that the occupation fraction factorizes as a product over the two merging halos.
To obtain the number density as a function of $M_\mathrm{bh}$ and $z$ alone, we integrate Eq.~\eqref{dn_informed} over mass ratio $q \equiv m_2/m_1 \in [0.1, 1]$, corresponding to a uniform integration in $Q \equiv 1/q \in [1,10]$. For the purposes of waveform generation and memory-jump amplitude calculations, we nonetheless restrict individual events to $q=1$, since the memory jump amplitude falls off sharply with increasing mass ratio, rendering the contribution of high mass-ratio events to the stochastic background negligible (see Fig.~\ref{fig:memorycharacteristics} or Sec.~\ref{sec:backgroundSNR}).

We follow the methods of Ref.~\cite{Langen:2024ygz} in computing this number density in practice. We discretize over a 3D grid in $z$, $M_\mathrm{bh}$, and $q$. To obtain values for the occupation fraction continuously over mass and redshift, we linearly interpolate the fitted values of $f_\text{bh}$ from $z=0.25$ to $z=3$, and hold $f_\text{bh}$ fixed at its $z=3$ value for higher redshifts. With this, the first three lines of Eq.~\ref{dn_informed} can be evaluated analytically at each grid point. For the halo mass function appearing in the fourth line, we use the package \texttt{HMFcalc} \cite{Murray:2013qza}, which returns $\mathrm{d}n_\mathrm{h}/\mathrm{d}\log_{10}M_\mathrm{h}$ at a given redshift per fixed comoving volume. The number density in Eq.~\ref{dn_informed} is therefore implicitly also per fixed comoving volume.

We plot the respective number density with respect to logarithmic mass and redshift of Model 2 in orange on Fig~\ref{fig:model_comparison}.
The total number of mergers per year can be obtained from the number density by integrating over the comoving volume shell and applying the light-cone correction \cite{Langen:2024ygz}:
\begin{align}
    \frac{dN}{dt} = \int_{0}^{z_\text{max}} dz \, \frac{dn}{dz} \, \frac{4\pi c \, D_L(z)^2}{(1+z)^2}.
\end{align}
Evaluating this numerically for 
the baseline/median parameters of Model 2 yields a total merger rate of approximately $216$ per year, consistent with Ref.~\cite{Langen:2024ygz}.

We emphasize a distinction between how mass ratio enters the merger-rate calculation and how it enters the memory-waveform calculation. The total merger rate and the number density of Eq.~\eqref{dn_informed} are computed by integrating over the full mass-ratio distribution $q \in [1,8]$ (or by other conventions [0.1,1]), as
prescribed by Ref.~\cite{Langen:2024ygz}, and are therefore unaffected by any
simplification applied downstream. 
When generating memory waveforms and computing the resulting stochastic background, however, we evaluate every sampled event at $q=1$ regardless of the mass ratio drawn from
Eq.~\eqref{dn_informed}. This substitution is motivated by the steep
suppression of the memory-jump amplitude with increasing mass ratio
(Sec.~\ref{sec:source_param_dep} and Fig.~\ref{fig:memorycharacteristics}): since $\Delta h^{\rm mem}$ falls off rapidly for $q>1$, evaluating all events at $q=1$ yields an optimistic upper bound on the memory contribution of each merger, and consequently on the amplitude of the stochastic background as
a whole. 
Additionally, we note that while $q$ does have a substantial effect on the characteristic rise time $\tau_c$, these mass ratio corrections are small relative to other sources of uncertainty in the background's duty-cycle statistics,
and do not materially affect its popcorn-noise character (see Sec.~\ref{sec:popcorn}). 
In conclusion, we adopt the $q=1$ approximation for both Model 1 and Model 2 to maintain a consistent, optimistic forecast across population models, while preserving the full $q$-dependence of the astrophysical merger rate
itself.

\section{Total Population Memory}\label{sec:stochasticbackground}
Now that we have discussed the memory waves from single events and the populations of SMBH mergers, let us turn our attention to many events throughout an observing run in LISA. 
We will first present our findings for predictions of a stochastic background, and follow with a discussion of interpretations of our results within LISA.

\subsection{Statistical Tools}

\begin{table}[htbp]
    \centering
    \begin{tabular}{lccc}
        \hline
        Parameter & Symbol & Range \\
        \hline
        Mass                  & $M/M_\odot$   & $[10^4-10^8]$\\
        Inclination           & $\theta$      & $[0,\pi/2]$  \\
        Redshift              & $z$           & $[0.1-10]$\\
        Luminosity Distance   & $D_L$         & Computed from $z$   \\
        Observing Time        & $T$           & 5 yrs \\ 
        \hline
    \end{tabular}
    \caption{Parameters for the Monte Carlo (MC) simulation. The mass distribution, $g(M)$, and the redshift distribution, $R(z)$, are not independent. They are derived from the joint astrophysical merger rate density, $dn / (dM dz)$, which defines the number of mergers per comoving volume, per unit mass, and per unit redshift.}
    \label{tab:bh_params}
\end{table}

The primary statistical tool used to characterize a stochastic gravitational-wave background (SGWB) is its power spectral density (PSD), denoted as $S_h(f)$. The single-sided
PSD is formally defined by the ensemble correlation of the Fourier-transformed strain, $\tilde{h}(f)$:
\[
    \langle \tilde{h}^*(f) \tilde{h}(f') \rangle = \frac{1}{2} \delta(f-f') S_h(f)
\]
Under a set of standard assumptions, the PSD is the \textit{only} defining feature needed to fully describe the background. These crucial assumptions are that the SGWB is a \textit{stationary}, \textit{Gaussian}, \textit{unpolarized}, and \textit{isotropic} process \cite{Zhang:2025rux, Allen:1997ad, Christensen:2018iqi, Romano:2016dpx}. In simple terms, this framework assumes:
\begin{itemize}
    \item The statistical properties of the source population do not change over the observation time (stationary).
    \item The background is composed of a large number of independent and overlapping signals, such that their sum approaches a Gaussian distribution by the Central Limit Theorem (Gaussian).
    \item The signal power is uniform across the sky (isotropic) and equal in both gravitational-wave polarizations (unpolarized).
\end{itemize}

\begin{figure}[t]
  \includegraphics[width=0.49\textwidth]{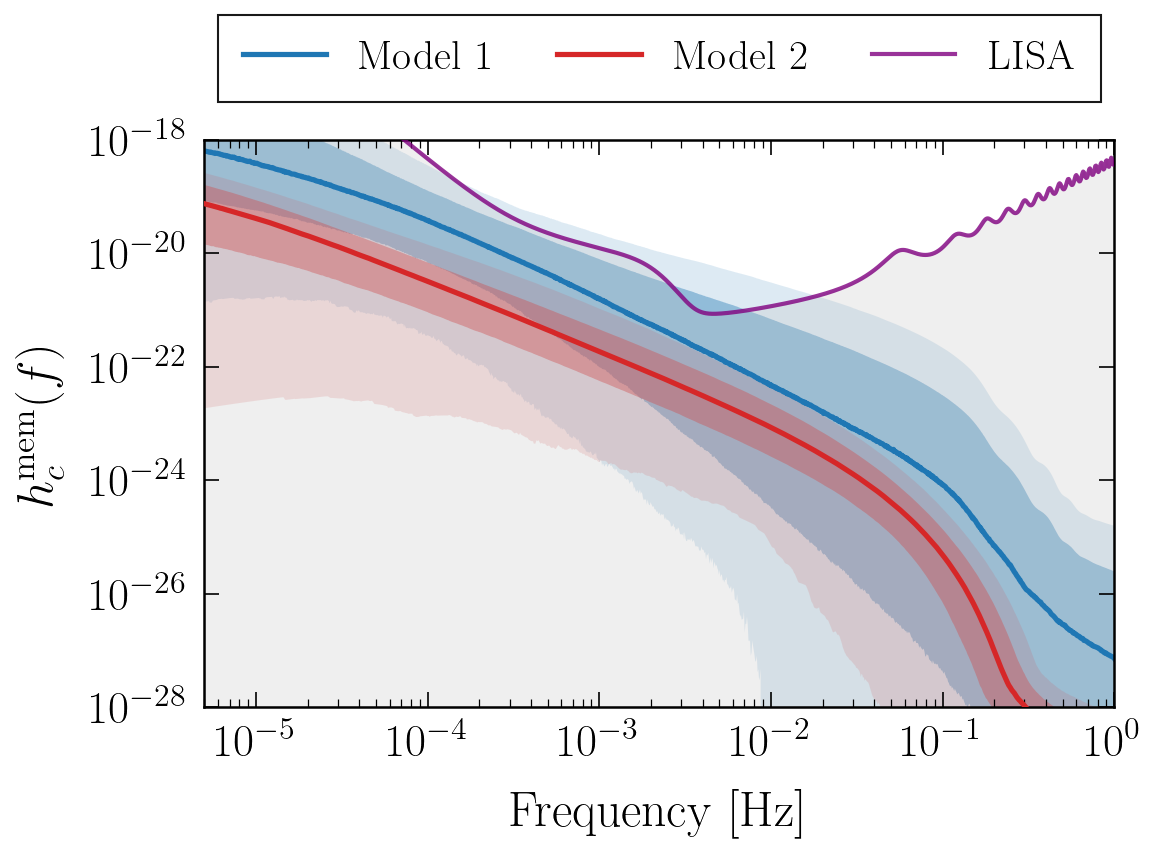}
  \caption{Our forecasts for the characteristic strain of the stochastic memory background from our MC runs. In blue is the Model 1 predictions, with light and dark bands representing 1$\sigma$ and 2$\sigma$ in model parameter space, respectively. Similarly in red, we have the predictions for Model 2 and its corresponding variance. We plot the LISA sensitivity curve for reference in purple. We find that out to $2\sigma$, in the Model 1 population predictions the characteristic strain of the SGWMB can range from factors of $10^3$ below to $10^1$ above the LISA noise curve. Conversely, Model 2 ranges from factors of $10^3$ to $10^1$ below the noise curve. We mitigate these faint estimates for Model 2 in Section~\ref{sec:popcorn}.}
  \label{fig:BackgroundMC}
\end{figure}

\begin{figure*}
  \centering
    \includegraphics[width=\textwidth]{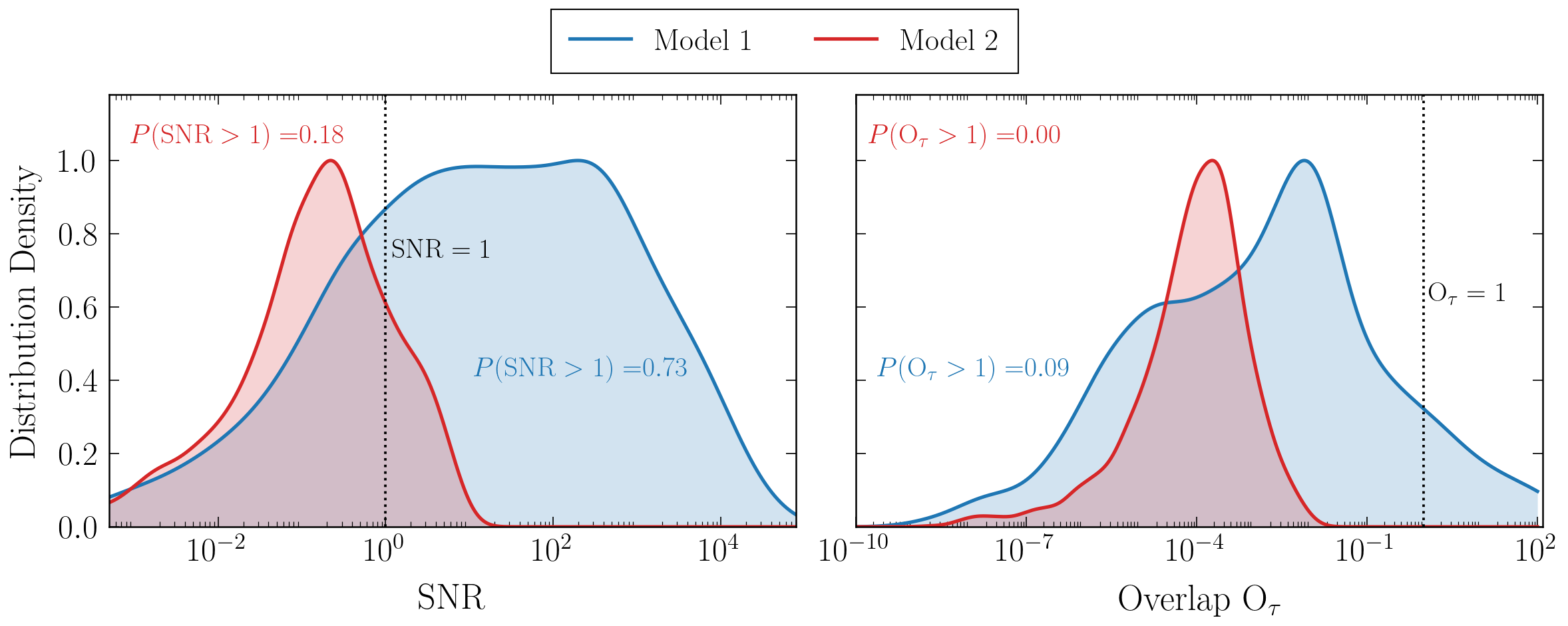}
  \caption{\textit{Left:} Distribution of
total SGWMB SNR across all realizations from Fig.~\ref{fig:BackgroundMC}, shown as peak-normalized KDEs.
Model 1 produces a detectable background in the majority of
realizations, while Model 2 is considerably fainter and only
occasionally exceeds an SNR of 1. The vertical dotted line at
$\mathrm{SNR}=1$ marks the rough detectability benchmark. \textit{Right}: Signal overlap density $\mathrm{O_\tau}$
(Eq.~\eqref{eq:overlap}, Sec.~\ref{sec:popcorn}) for individual realizations of the SGWMB, shown as peak-normalized KDEs across the Monte Carlo ensemble. Model 1 (blue) shows substantially larger scatter than Model 2 (red), which is nearly always resolvable into individual burst events. The vertical dotted line at $\mathrm{O_\tau}=1$
marks the rough threshold separating popcorn ($\mathrm{O_\tau}<1$) and
continuous ($\mathrm{O_\tau}>1$) regimes.}
  \label{fig:distribution_density}
\end{figure*}

Similarly to the idea of an SGWB, many GW events over time are predicted to also leave behind memory imprints, leading to the notion of a stochastic GW memory background (SGWMB) if the individual memory events cannot be resolved on their own. In any situation where the background is composed of a high number of discrete, step-like memory events, the accumulated strain can be statistically described as a random walk. More specifically, if the increments of this walk are assumed to be independent and normally distributed, an idealization that may hold for a large merger population of sources, the process mathematically resembles Brownian motion. In this idealized framework, the evolution of the strain is essentially fully characterized by a single parameter: a diffusion coefficient, $D$. In the context of the gravitational waves, this diffusion coefficient can be used when approximating individual time domain memory signals as a Heaviside step function, a concept explored in \cite{Zhao:2021zlr, Zhang:2025rux}.

However, for a background generated by the cumulative effect of gravitational-wave memory, there is a strong expectation that the standard SGWB assumptions will not hold in light of the population of SMBH mergers. Unlike the oscillatory part of the signal, the memory background is composed of fewer, more distinct ``step-like'' events from the most massive binary mergers. This can lead to a ``popcorn'' or shot-noise character, violating the Gaussianity assumption. Furthermore, the low rate of these dominant events may challenge the assumption of stationarity.

Despite these significant caveats, calculating an effective $S_h(f)$ for the memory background remains a valuable exercise. It provides a standard measure of the background's power, allowing for direct comparison with other astrophysical and cosmological backgrounds and with the sensitivity curves of LISA. Therefore, we proceed with this calculation, keeping in mind that $S_h(f)$ provides an incomplete, though highly useful, description of the memory background's true nature \cite{Mingarelli:2019mvk}.

Armed with this, we choose to model each 
type of memory event, 
as determined from their SMBH mass configuration, as a random Poisson process, with randomly chosen parameters listed in Table~\ref{tab:bh_params}. 
Given a signal parameter $\theta$, if the rate is $R(\theta) d\theta$, and the frequency-domain signal is $\tilde h^{\rm mem}(\theta,f)$, then the corresponding PSD is given by 
\begin{equation}
S_h (f) = \int     R(\theta) |(\tilde h^{\rm mem})^2 (\theta,f)|d\theta 
 \end{equation}
 In our analysis, we construct a background by creating Monte Carlo samples of independent mergers that take place within observation time $T$ and estimate the energy density spectrum directly.  More specifically, for a duration $T$, by choosing a particular population model discussed in Sec.~\ref{sec:population} and illustrated in Fig.~\ref{fig:model_comparison}, we can obtain a Monte Carlo sample population containing $N$ events with parameters  $\theta_j$, $j=1,\ldots, N$.  From this collection of events we can estimate the PSD as 
 \begin{equation}
S_h[\{\theta_j\}] = \frac{1}{T}\sum_{j=1}^N     |(\tilde h^{\rm mem})^2(\theta_j,f)|
\label{eq:Sn_calc}
 \end{equation}

After computing $S_h$, we can calculate the characteristic strain \cite{Zhao:2021,Zhao:2021zlr, Boybeyi:2024aax}:
\begin{align}
    h_c^{\text{mem}}(f) = \sqrt{2f S_h(f)} \label{eq:hc}
\end{align}

Under the assumption that the background is isotropic, the optimal signal-to-noise ratio (SNR) after an observation time $T$ in LISA can be uncovered for our known PSD $S_h$ and detector overlap function $\Gamma_{12}$. We approximate this function for LISA to be 1, drawing from Fig. 36 of~\cite{Romano:2016dpx}. Then, the SNR
is given by Eq. 6.34 of~\cite{Romano:2016dpx}:
\begin{align}
    \text{SNR} = \sqrt{T} \left( \int_{0}^\infty df \cfrac{\Gamma_{12}^2(f) S_h^2(f)}{\mathcal{R}^2(f)S_n^2(f)} \right)^{1/2}
    \label{eq:snr}
\end{align}
where we have used the relation $S_n(f) = P_n(f) / \mathcal{R}(f)$. $P_n(f)$ is the power spectral density of the LISA detector noise, and $\mathcal{R}(f)$ is the sky and polarization averaged signal response function. 
Ref. \cite{Robson:2019qvq} well approximates this function as 
\begin{align}
    \mathcal{R}(f) = \frac{3}{10} \cfrac{1}{(1 + 0.6(f/f_*)^2)}
    \label{eq:response}
\end{align}
with $f_* = 19.09$ mHz, known as the transfer frequency~\cite{Prince:2002hp}. 
We note that in the low frequency limit, many references approximate $\mathcal{R}=3/20$; the 3/10 value quoted in Eq.~\ref{eq:response}, and used in~\cite{Robson:2019qvq} comes from summing over two independent low-frequency data channels.
$f_*$ sets the scale at which finite-arm effects begin to suppress the response function. $\mathcal{R}(f)$ applies best below $f_*$, where LISA has two independent data channels; above $f_*$ a third channel opens, and the three-channel expressions of Ref.~\cite{Prince:2002hp} become more appropriate~\cite{Robson:2019qvq}.
For comparison, NR memory waveforms begin to roll off above approximately $f \sim 1/(70M_z)$, which for the masses considered here lie comparable to or below $f_*$. The memory content is therefore mainly concentrated in the band where the two-channel response applies.

\begin{figure}
  \centering
  \includegraphics[width=0.49\textwidth]{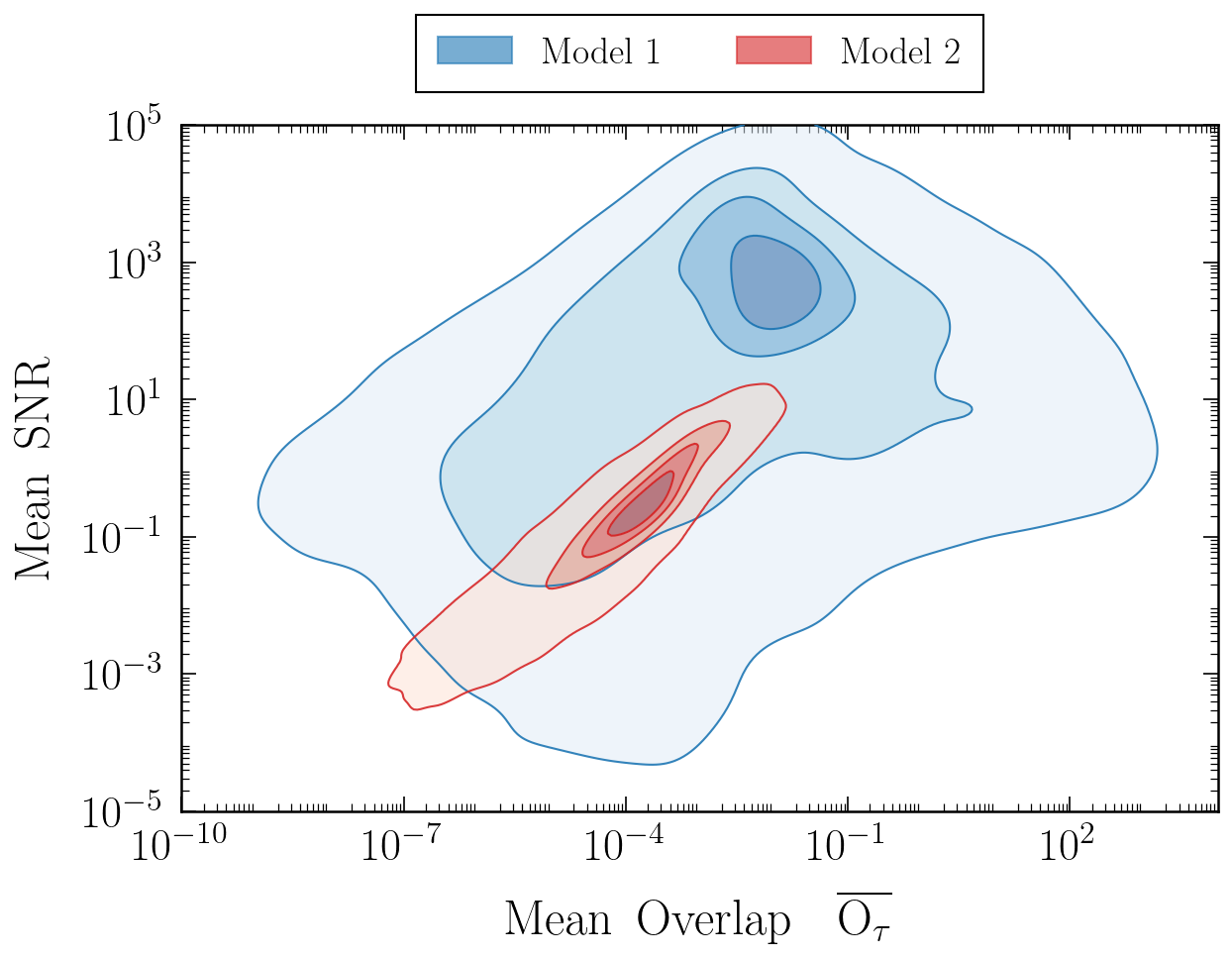}
  \caption{
  Joint distribution of mean SGWMB SNR and mean signal overlap $\overline{\mathrm{O_\tau}}$ across the population hyperparameter space, for Model 1 (blue) and Model 2 (red). Each point corresponds to a fixed draw of population hyperparameters; contours reflect the density of these draws under each model's prior, with darker contours indicating higher density.}
  \label{fig:snr_vs_numMerger}
\end{figure}

\subsection{Background SNR and Dependence on Population Models}
\label{sec:backgroundSNR}

Fig.~\ref{fig:BackgroundMC} shows our forecasts of the SGWMB with characteristic strains drawn out to $2\sigma$ in our population model parameter space.
We highlight the variance for both $1\sigma$ and $2\sigma$ is much larger for our Model 1 population than for Model 2, as expected. Using the results of Fig.~\ref{fig:BackgroundMC}, 
in the left plot of Fig.~\ref{fig:distribution_density} we plot the normalized probability density of the SGWMB SNR for both populations overlaid. We draw attention to the line of SNR = 1, where we roughly assume that the SGWMB transpires from being subdominant. For Model 1, the probability that the SNR exceeds one is more likely than not, at 73\%, while for Model 2 it is forecasted at 18\%. 
The median SNR differs between the two populations, with Model 1 yielding $\sim 16$ and Model 2 yielding $\sim 0.179$, reflecting the substantially more optimistic source counts admitted by Model 1. We report that the $1\sigma$ and $2\sigma$ SNR ranges of the background are [0.218, 768] and [0.00149, 9605], respectively, for our Model 1 population. Similarly, we find the $1\sigma$ and $2\sigma$ bands correspond to [0.0174, 1.190] and [0.000925, 4.79] for our Model 2 population.

We note that within the $1\sigma$ range for the Model 2 population, the SGWMB SNR is of order $\mathcal{O}(1)$ or less, indicating that this population does not constitute a significant stochastic foreground for LISA. 
This is not the case for Model 1's population, whose $1\sigma$ upper bound extends to order $\mathcal{O}(10^2)$. This substantially greater dispersion is a direct consequence of the permissive nature of Model 1's population, which, by construction, 
imposes no astrophysical constraints on the merger rate or mass distribution, thereby admitting highly optimistic realizations of the source population with correspondingly large SGWMB amplitudes. The Model 2 population, by contrast, is anchored to observational constraints on the merger rate and mass function, suppressing such extreme realizations and yielding a significantly narrower SNR distribution, with the $2\sigma$ upper bound remaining at order $\mathcal{O}(10^1)$ compared to $\mathcal{O}(10^5)$ for Model 1, as illustrated in Fig.~\ref{fig:snr_vs_numMerger}.

\subsection{Popcorn nature of the background }

We quantify the statistical nature of the background by calculating the signal overlap density (or shorthandedly overlap), defined as 
\begin{equation}
    \mathrm{O_\tau} = \frac{1}{T} \sum_{i} \tau_{c,i}
    \label{eq:overlap}
\end{equation}
for each realization in our Monte Carlo simulation. Namely, we sum over the dominant portion of the memory signal for each event in our forecast, and then divide by the total five-year observing period: $T$. This quantity serves as a proxy for the validity of the Central Limit Theorem assumption. We define that $\mathrm{O}_\tau = 1$ corresponds to
the situation where, on average, there is one memory signal active at any moment the time domain.\footnote{One could imagine this case consisting of memory signals that are lined up ``back to back'' across your entire observing time. On the contrary, since the overlap is an average, you could have a single window where a large amount of mergers happen at the same time and cancel out the observation-time-average and produce an identical overlap. We assume the events are temporally distinct enough that the overlap is not probing situations that are so unique as the latter, where it would effectively lose all meaning.}
Then,
a value $\gg 1$ implies a continuous, Gaussian background, whereas a value $\ll 1$ indicates a ``popcorn'' regime of discrete, non-overlapping bursts. 
Because the arrival times are random, such a realization will generically contain both overlapping bursts and quiet gaps. Then, more usefully, $\mathrm{O_\tau}$ may be read as the expected number of signals active at any given moment: for $\mathrm{O_\tau}\ll1$ any two bursts are unlikely to coincide, while for $\mathrm{O_\tau}\gg1$ many signals overlap at every instant, as required for the Central Limit Theorem to apply. We therefore adopt $\mathrm{O_\tau}=1$ as the demarcating value between the two regimes, since for $\mathrm{O_\tau}>1$ the ensemble statistics \textit{begin} to reflect consistent signal overlap.

As illustrated in the right plot of Fig.~\ref{fig:distribution_density}, the vast majority of realizations for both population models result in a low overlap. Qualitatively, we find the median overlap for model 1 to be 
$\approx 2\times 10^{-3}$ with $1\sigma$ and $2\sigma$ ranges at 
$[7.25\times 10^{-6}, 1.59\times 10^{-1}]$
and 
$[8.03\times 10^{-8}, 2.55\times 10^{1}]$, respectively.
Likewise, for Model 2, we find the median to be
$\approx 10^{-4}$ with $1\sigma$ and $2\sigma$ ranges at 
$[8.82\times 10^{-6}, 5.72\times 10^{-4}]$
and 
$[1.64\times 10^{-7}, 3.10\times 10^{-3}]$, respectively. This places both populations deep in the sparse-overlap regime, for which a Gaussian approximation is unlikely to be adequate. Notably, the Model 2 population is non-Gaussian, with a maximum overlap of order $\mathcal{O}(0.01)$.\footnote{
Since $\mathrm{O_\tau}$ is linear in $\tau_c$, the effect of this simplification can be bounded directly. For nonspinning binaries, $\tau_c$ at $q=8$ is roughly $4$--$5$ times larger than at $q=1$, so a population composed entirely of $q=8$ events would rescale the windowed overlap to a maximum of $\mathrm{O_\tau} \sim 0.1$, notably still far below 1. A more realistic sampling of the SMBH population with varying $q$ falls between these values. Therefore our popcorn classification is robust, even with the equal-mass assumption.
}
Conversely, while Model 1's population exhibits larger variance, permitting rare realizations where the background approaches Gaussian statistics, the median and mode of the distribution remain well within the intermittent, non-Gaussian regime. 

We further examine this contrast through the joint behavior of the SGWMB mean SNR and mean overlap $\overline{\mathrm{O_\tau}}$
across the population hyperparameter space, shown in Fig.~\ref{fig:snr_vs_numMerger}. Rather than propagating a single Monte Carlo draw of the population directly into these quantities, we fix a draw of the population hyperparameters and run an inner Monte Carlo over stochastic realizations at that point. Then, we compute $\overline{\mathrm{SNR}}$ and 
$\overline{\mathrm{O_\tau}}$ for each run before repeating this across a dense sampling of the hyperparameter space. This isolates the distribution of the population models themselves from the additional scatter introduced by individual stochastic realizations, giving a more direct probe of how $\overline{\mathrm{SNR}}$ and $\overline{\mathrm{O_\tau}}$ vary across each model's parameter space. 
We find that
Model 1 occupies a broad region of this space, with a high-density lobe near $\overline{\mathrm{O_\tau}} \sim 10^{-3}\text{--}10^{-2}$ and $\overline{\mathrm{SNR}} \sim 10^{2}\text{--}10^{3}$, and substantial probability extending to low SNR. Model 2, by contrast, concentrates in a narrow, 
elongated region at lower overlaps ($\overline{\mathrm{O_\tau}} \sim 10^{-7}\text{--}10^{-4}$) and moderate SNR ($\overline{\mathrm{SNR}} \sim 10^{-2}\text{--}10^{0}$). This compactness, again a signature of Model 2's galactic anchoring, confines it to a tight space in the SNR--overlap plane, whereas Model 1's relatively more unconstrained parametrization spreads across orders of magnitude in both. 

Critically, Model 2's overlap distribution shows essentially no support for temporally overlapping signals over a full five-year LISA observing run. Since we take Model 2 to be the more conservative of the two population models, we conclude that a stationary Gaussian-noise model is unlikely to capture the statistics of the SGWMB; rather, a collection of discrete, or ``popcorn'', transients is likely to be more accurate. This popcorn nature has a direct implication for SNR estimation: integrating coherently over the full mission lifetime $T_{\rm obs}$ dilutes the accumulated SNR with time during which no memory transient
is active.

\subsection{Windowing Treatment of the Background}
\label{sec:popcorn}

In this subsection we address the following question: how much of the apparent weakness of the effective five-year PSD comes simply from treating a sparse population of transients as if it were continuously present? 
By construction, Eq.~\eqref{eq:Sn_calc} spreads the power of every burst across the full observing time $T$. Therefore, for a popcorn background, the resulting power $S_h(f)$ and overlap function $\rm{O}_\tau$ are diluted by the long stretches of data that contain no memory signal at all.
We restrict our attention to Model 2, which exhibits both a faint and  popcorn nature across all realizations of the SGWMB.
To test if these predictions are an artifact of the observing time rather than a robust feature of the population, we restrict our attention to an ``active'' window around each merger. 
We highlight that this is a \textit{diagnostic} for isolating the dilution, internal to our forecasts, and not a data-analysis prescription for a real search in LISA.

Because the memory strain accumulates almost entirely within the jump, an appropriately chosen window captures all the strain needed to study memory. We set each window $w_i = t_{2,i} - t_{1,i}$ by the times on either side of the burst where the news amplitude $|\dot h_i^{\rm mem}(t)|$ falls below 10\% of its peak. Since the threshold is applied relative to the peak, the construction is agnostic to the mass scaling of the waveform and requires no prior knowledge of $\tau_c$ or of exactly when the merger occurs. 
Then, the active observing time is the duration of the union of these windows, $\bigcup_i [t_{1,i}, t_{2,i}]$, accounting for overlap between neighboring events, and is used in place of the full five-year baseline in Eq.~\eqref{eq:overlap}. 
\footnote{We note that our windows, however, are still built from the noiseless waveforms. The most agnostic approach would apply a global threshold on the news amplitude to noisy realizations of the full data stream, which would contribute staying agnostic of the exact shape of the memory waveform.}

\begin{figure}
    \centering
    \includegraphics[width=\linewidth]{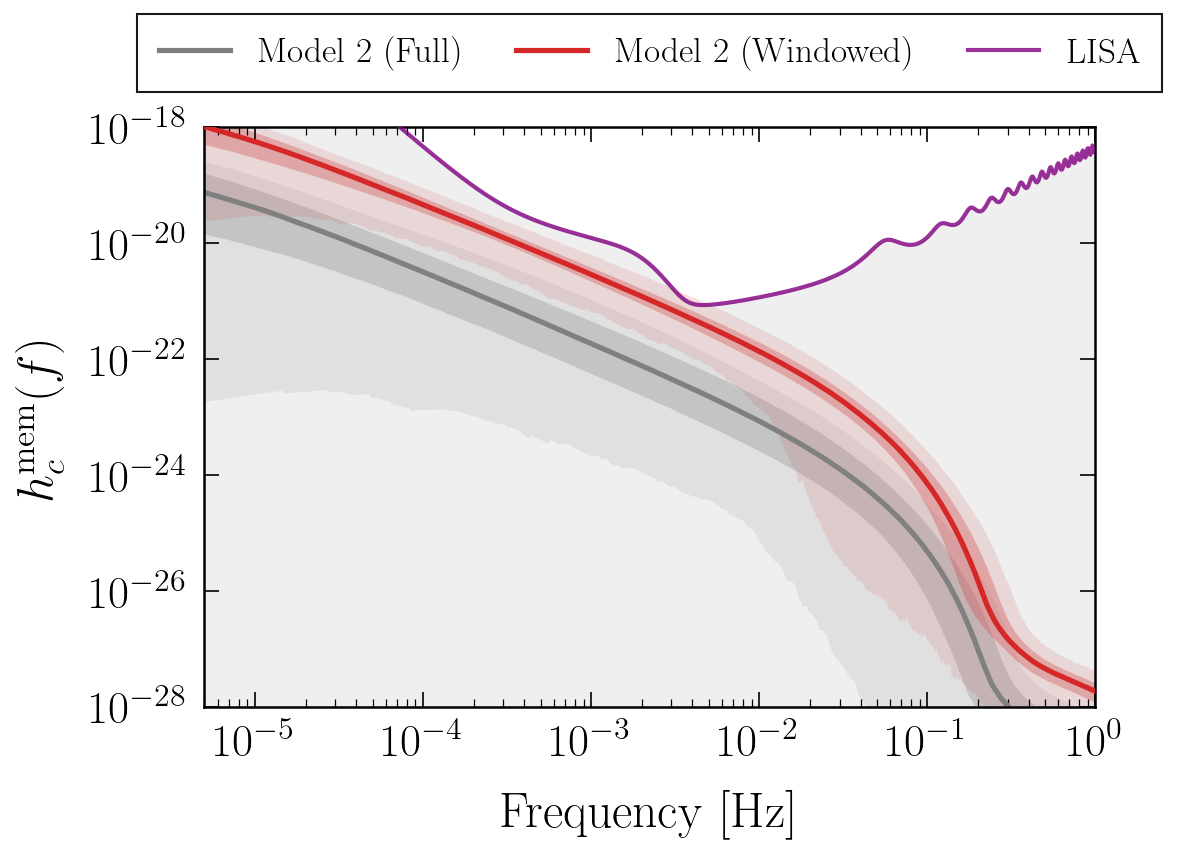}
    \caption{SGWMB forecasts for our Model 2 population while mitigating the
    ``popcorn'' Gaussianity problems presented in Section~\ref{sec:popcorn} with our windowed approach. Similar to Figure~\ref{fig:BackgroundMC}, the lighter and darker shading corresponds to one sigma and two sigma likelihood draws from our population parameter space. 
    We present the characteristic strain median and variance from Figure~\ref{fig:BackgroundMC} in gray for reference, and the results from the windowed approach in red. 
    }
    \label{fig:window_informed_mc}
\end{figure}

\begin{figure*}[t]
    \centering
    \includegraphics[width=\textwidth]{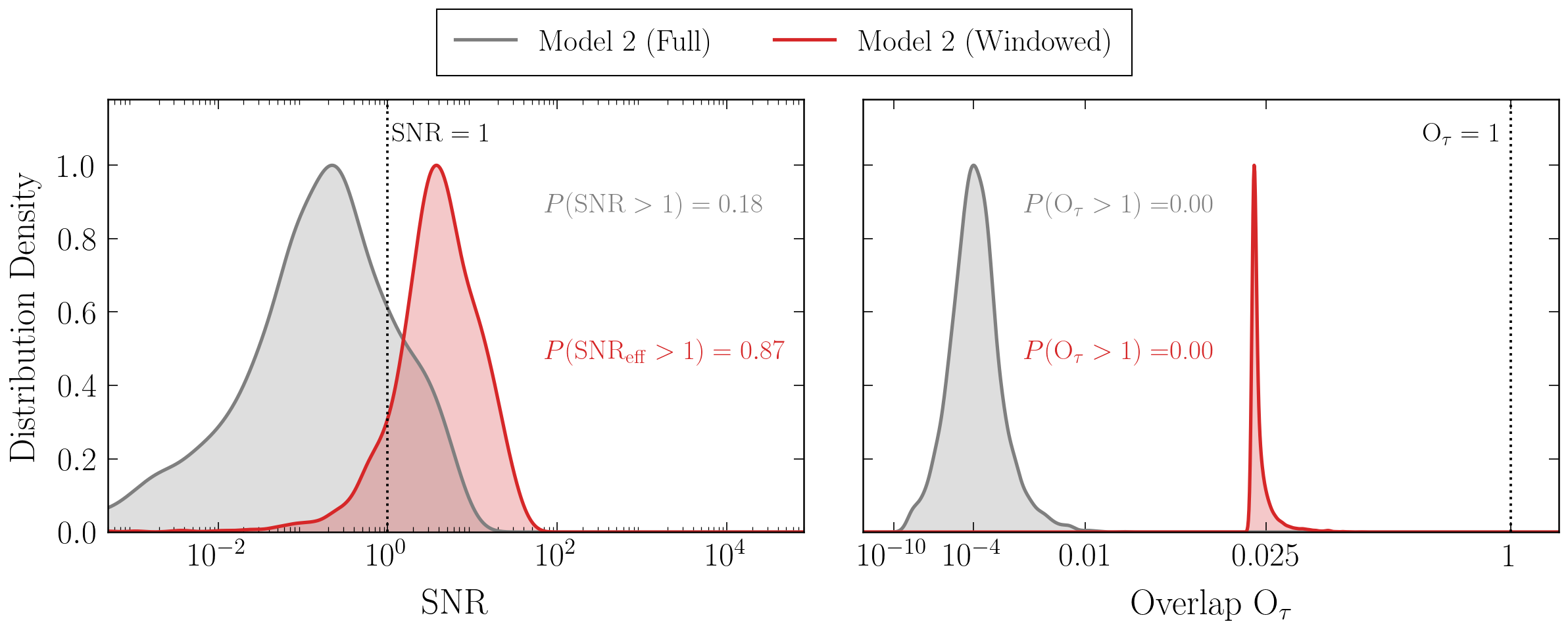}
    \caption{Our SNR and overlap distribution densities for our windowed approach (red) and Model 2 population using the results from Fig.~\ref{fig:window_informed_mc}. As in Fig.~\ref{fig:window_informed_mc}, we plot the non-windowed distributions from Fig.~\ref{fig:distribution_density} for reference in gray. 
    Similar to Fig.~\ref{fig:distribution_density}, we denote characteristic thresholds of SNR $= 1$ and $\mathrm{O_\tau} = 1$ with a dotted red line. We find that approximately 87\% of our realizations of the SGWMB lie above the SNR threshold, while still 0\% in the overlap lie above 1, although our windowed approach provides at least a factor of 10 improvement. Likewise, the overlap features a factor of 100 improvement, albeit there is still a vanishingly small chance of realizations exceeding an overlap of 1.}
    \label{fig:window_informed_densities}
\end{figure*}

The reason this collection of memory events nonetheless remains incoherent, rather than reducible to event-by-event resolved detections, is that our statistics do not condition on the parameters of individual bursts. 
This is appropriate because knowing \textit{when} a merger occurs fixes neither the sign nor the amplitude of its memory, since both depend sensitively on the binary's configuration and orientation, which we treat as unconstrained across the population.
Conversely, if the binary parameters and sign of the jump are known or can be accurately marginalized over, one could stack these memory bursts coherently. The fully coherent and incoherent treatments can be viewed as the two limiting assumptions for combining multiple events, in which the per-event parameters are either identical or entirely unrelated. 
Both are limits of a hierarchical framework, and it has been shown that
selecting either of these limits will decrease detection ability~\cite{Zimmerman:2019wzo}.

We repeat our analysis of Figs.~\ref{fig:BackgroundMC} and
\ref{fig:distribution_density} using this windowed approach for population Model 2, which had initially yielded the faintest
forecasts of both effective active-time SNR ($\mathrm{SNR_{active}}$) and overlap. Fig.~\ref{fig:window_informed_mc}
presents the resulting characteristic strain alongside the full approach for reference, and Fig.~\ref{fig:window_informed_densities} presents the corresponding $\mathrm{SNR_{active}}$ and overlap densities. Under full integration, the median SNR is 0.18, with only 18\% of realizations
exceeding our ``detection threshold'' of $\mathrm{SNR}=1$; restricting
the integration
window to periods of active signal raises the median $\mathrm{SNR_{active}}$ to 3.915 and increases the detectable fraction to 87\%. Furthermore, we see a constriction in the width of the windowed $1\sigma$ and $2\sigma$ bands in $\mathrm{SNR_{active}}$: [1.404, 1.478] and [0.235, 2.647] respectively. 
The overlap shows a comparable narrowing. Rather than the broad, faint distribution seen under full integration, the windowed overlap
clusters tightly around 0.023. Quantitatively, the overlap is at a median of 0.0246 with $1\sigma$ and $2\sigma$ ranges at [0.0246, 0.0251] and [0.0246, 0.0263] respectively.  

In summary, we find that when the observing time is fixed at $T_{\rm obs}$, a substantial share of the variance in characteristic strain, 
SNR, 
and overlap across realizations arises from the stochastic number of mergers $N$ that occur. We focus that
the specific value of $\mathrm{O}_\tau$ is, as noted above, an artifact of the equal-mass and nonspinning simplification. 
However we urge that its magnitude relative to $\mathrm{O}_\tau=1$ is a feature that will persist in a more realistic population, as it is purely a statement about how sparsely bursts populate the observing time.

\section{Discussion}
In this paper we have presented forecasts for the stochastic background of GW memory within the LISA detector. 
We have uniquely characterized the cumulative signal from a population of SMBH mergers with three distinct features from previous studies of memory backgrounds. 
First, rather than approximating each memory burst as a Heaviside step function, we modeled it with the \texttt{NRHybSur3dq8-CCE} surrogate waveform, which resolves the finite rise time of the signal and the high-frequency cutoff of the memory spectrum. Second, we chose two astrophysically distinct SMBH population models and extensively analyzed the differences that choice of population holds for any model of the 
SGWMB. Third, we characterized not only the average power of the background but also its time-domain statistics, quantifying its departure from the Gaussian, stationary idealization.

This statistical nature of the SGWMB has significant implications for LISA data analysis. 
The SGWMB's primary relevance takes the form of
an astrophysical confusion noise that must be accurately characterized and subtracted. Normal stochastic backgrounds take on the assumptions idealized in Sec. \ref{sec:stochasticbackground}, and can be characterized to high accuracy in the present.
However, we have shown that the SGWMB, with
its sparse event rate, causes the Central Limit Theorem (CLT) assumption to fail. This results in the background emerging as a popcorn collection of discrete bursts in the time domain. We quantify this with the overlap $\rm{O}_\tau$, which is typically at values $\ll 1$ across both population models (median $\sim \mathcal{O}(10^{-3})$ for Model 1, $\mathcal{O}(10^{-4})$ for Model 2, and $0.025$ for Model 2 with the windowed approach). 
Due to this agreement despite population model variance, our results indicate that individual memory bursts should be expected to remain temporally distinct in LISA.  
This will form a background that is a fundamentally different statistical object than one created by a continuous, stationary process, and we urge its proper characterization will be important for the LISA global fit.

The LISA global fit decomposes its data stream into resolvable sources, parameterized stochastic components, and instrumental noise~\cite{Rosati:2024lcs, Katz:2024oqg}. Our results locate the memory background within this 
framework in two ways. For the loud, individually resolved SMBH mergers, the merger time will be known to high precision, so the accompanying memory burst could be searched for coherently and ultimately be absorbed into the waveform models of the resolved events themselves~\cite{Inchauspe:2024}. 
The memory of the remaining, unresolved population instead enters the data as an intermittent, non-Gaussian residual. By the overlap statistics of Sec.~\ref{sec:popcorn}, this residual is unlikely to be captured by the stationary, Gaussian stochastic components that the global fit would conventionally use. Cross-correlation searches built on the 
stationary and Gaussian assumption may thus lose sensitivity to this component and risk misattributing its power, 
whereas search frameworks designed for intermittent, non-Gaussian backgrounds~\cite{Thrane:2013kb, Cornish:2015pda, Drasco:2002yd} provide a more natural starting point. Left unmodeled, we urge that such a residual could bias the recovery of other stochastic signals in the LISA band, including the galactic confusion foreground and possible cosmological backgrounds.

In conclusion, we expect the SGWMB in LISA to be popcorn in statistical nature, limited in individual event subtractability, 
and strongly SMBH population model-dependent. Its detectability is not assured, and the background reaches detectable SNR in the majority of Model 1 realizations, while under a more conservative Model 2 it most likely remains undetectable when integrated over the full mission. 
We highlight that the large variance in our predictions between population models exposes the limitations of our current astrophysical knowledge about SMBHs, which
underscore the need to better understand their true population before the LISA mission begins observations. Improving constraints on merger rates and mass distributions, perhaps by incorporating data from pulsar timing arrays, is critical for refining forecasts. Although astrophysical uncertainty complicates predictions for LISA, the mission itself is set to provide accurate measurements on the SMBH population. We look forward to when LISA's conclusion on the SMBH population replaces the broad model variances presented here, transforming our understanding of the formation of galaxies and their co-evolution with black holes.

\acknowledgments
We would like to thank Dr. Kwinten Fransen for early collaboration on this work. We would like to thank Dr. Kathryn Zurek for insightful feedback on this work. 
J.~B., A.~L., and Y.~C.'s research is supported by the Simons Foundation (Award No. 568762), the Brinson Foundation, and the National Science Foundation (via Grants No. PHY-2011961 and No. PHY-2011968).
A.L. is grateful for support from the Fannie and John Hertz Foundation in the form of a Hertz Fellowship. J.B. is grateful for the support from the Caltech VURP fellowship. 

\bibliography{main}

\end{document}